\documentclass{aa}  

\usepackage{graphicx}
\usepackage{txfonts}
\usepackage[colorlinks=true, allcolors=blue]{hyperref}
\usepackage{stfloats}
\usepackage{booktabs} 
\usepackage{hyperref}
\usepackage{natbib}

\begin{document}

   \title{The effect of measurement uncertainties on the\\ inferred
   stability of planes of satellite galaxies}
   \titlerunning{Effects of uncertainties on inferred satellite plane stability}
   \authorrunning{Kumar et al.}

   \author{Prem Kumar
          \inst{1,2},
          Marcel S. Pawlowski\inst{2},
          Kosuke Jamie Kanehisa \inst{2,3},
          Pengfei Li \inst{4},
          Mariana P. J\'{u}lio \inst{2,3}
          \and 
          Salvatore Taibi \inst{2,5}
          }

        \institute{Institut f\"ur Astrophysik (IfA), Universit\"at Wien,  T\"urkenschanzstraße 17, 1180 Wien\\
        \email{prem.kumar@univie.ac.at}
        \and
        Leibniz-Institut f\"ur Astrophysik Potsdam (AIP), An der Sternwarte 16, 14482 Potsdam, Germany\\
        \email{mpawlowski@aip.de}
        \and
        Institut f\"ur Physik und Astronomie Universit\"at Potsdam, Karl-Liebknecht-Straße 24/25, 14476 Potsdam, Germany
        \and
        School of Astronomy and Space Science, Nanjing University, Nanjing, Jiangsu 210023, China
        \and
        Institute of Physics, Laboratory of Astrophysics, Ecole Polytechnique Fédérale de Lausanne (EPFL), Observatoire de Sauverny, CH-1290 Versoix, Switzerland
             }

   \date{Received September 15, 1996; accepted March 16, 1997}

 \abstract
   {Observations have revealed that the Milky Way, Andromeda, Centaurus A, and possibly other galaxies host spatially thin and kinematically coherent planes of satellites. Such structures are highly improbable within the standard $\Lambda$CDM cosmological model, and the dynamical stability of these planes has long been debated. Accurately determining their stability requires a thorough understanding of orbital parameters such as proper motion, distance, and line-of-sight velocity, in addition to as the gravitational potential of the host galaxy. However, many of these parameters remain poorly constrained, leading to significant uncertainties in analyses.}
   {This study explores the impact of measurement errors in the proper motions and distances of the satellite galaxies and in
the adopted host halo mass on the inferred stability of satellite planes in Milky Way-like potentials.}
   {We simulated mock-observed test satellite galaxies orbiting a host galaxy by adding various degrees and types of observational
errors, and then backward-integrated the orbits. We analyzed trends and correlations between the initial conditions and the uncertainties applied on the inferred orbital stability of the satellite systems. We also considered the effects of adopting incorrect potentials and the impact of different orbital eccentricities.}
   {Uncertainties in proper motions lead to an apparent widening of an intrinsically stable satellite plane, with its width increasing linearly with the uncertainties in the adopted proper motion. Even uncertainties at the level of Gaia systematics strongly affect the plane's inferred past width. Moreover, the potential with a low halo mass has a significant impact on the stability of these planes, whereas the remaining two host models show similar effects. Uncertainties in satellite distance also contribute noticeably to the inferred instability.}
   {}


   \maketitle

\section{Introduction}
A century ago, astronomers were aware of only two prominent satellite galaxies orbiting the Milky Way, the Large Magellanic Cloud (LMC) and the Small Magellanic Cloud (SMC), a view that had persisted for centuries \citep{von1859cosmos}. This census doubled when \cite{shapley1938two} used photographic plates to discover two additional dwarfs, Sculptor and Fornax. With the discovery of Sagittarius in 1994, the number of known Milky Way satellite galaxies reached 11 \citep{ibata1994dwarf}, and has since increased substantially due to advances in observational technology, the deployment of larger telescopes, and systematic sky surveys. Currently, approximately 60 dwarf galaxies have been discovered around the Milky Way \citep{simon2019faintest}. Notably, these bright satellite galaxies tend to lie along a great circle known as the “Magellanic Plane” \citep{lynden1976dwarf}. Observations show that Milky Way satellite galaxies are highly anisotropic and kinematically correlated, with most located in a thin plane roughly perpendicular to the disk of the Milky Way \citep{kroupa2005great, metz2008orbital}. The most luminous among these galaxies appear to align within a thin, disk-like structure known as the Vast Polar Structure (VPOS; \citealt{pawlowski2012,taibi2024}). The VPOS exhibits a root mean square (rms) thickness of 20-30 kpc,  a radius of about 250 kpc, and an axis ratio $c/a$ of 0.18 -- 0.30 \citep{pawlowski2018planes, pawlowski2021phase}. 

Similarly, a comparable thin and anisotropic distribution of satellite galaxies is also found around Andromeda, referred to as the Great Plane of Andromeda (GPoA) \citep{ibata2013vast, conn2013three}, and around Centaurus A, known as the Centaurus A Satellite Plane (CASP) \citep{tully2015two, muller2018whirling,kanehisa2023a}. These structures are characterized by an axis ratio of $c/a$ = 0.1 \citep{pawlowski2013dwarf} and a $c/a$ ratio = 0.2 \citep{tully2015two}, respectively. Both the GPoA and CASP planes exhibit significant correlations in the line-of-sight velocities of their satellite members when observed edge-on \citep{ibata2013vast, muller2021coherent}. This suggests that satellites corotate, with satellites on one side of the host predominantly receding and those on the other side approaching relative to the host. This rotating pattern around its host galaxy, similar to the VPOS, supports the idea that both the GPoA and CASP constitute a nonrandom, coherent structures. Moreover, the first proper motion measurements for three on-plane satellite galaxies of M31 indicate that these galaxies are consistent with co-orbiting along the spatially identified GPoA \citep{sohn2020,pawlowskisohn2021,casettidinescu2024}. A similar flattened structure with coherent line-of-sight velocities was discovered around the galaxy NGC\,4490 and is found to be rare in cosmological simulations \citep{karachentsev2024, pawlowski2024}.

\label{Sect:Method}

\begin{figure*}[t]
\centering
   \includegraphics[width=15    cm]{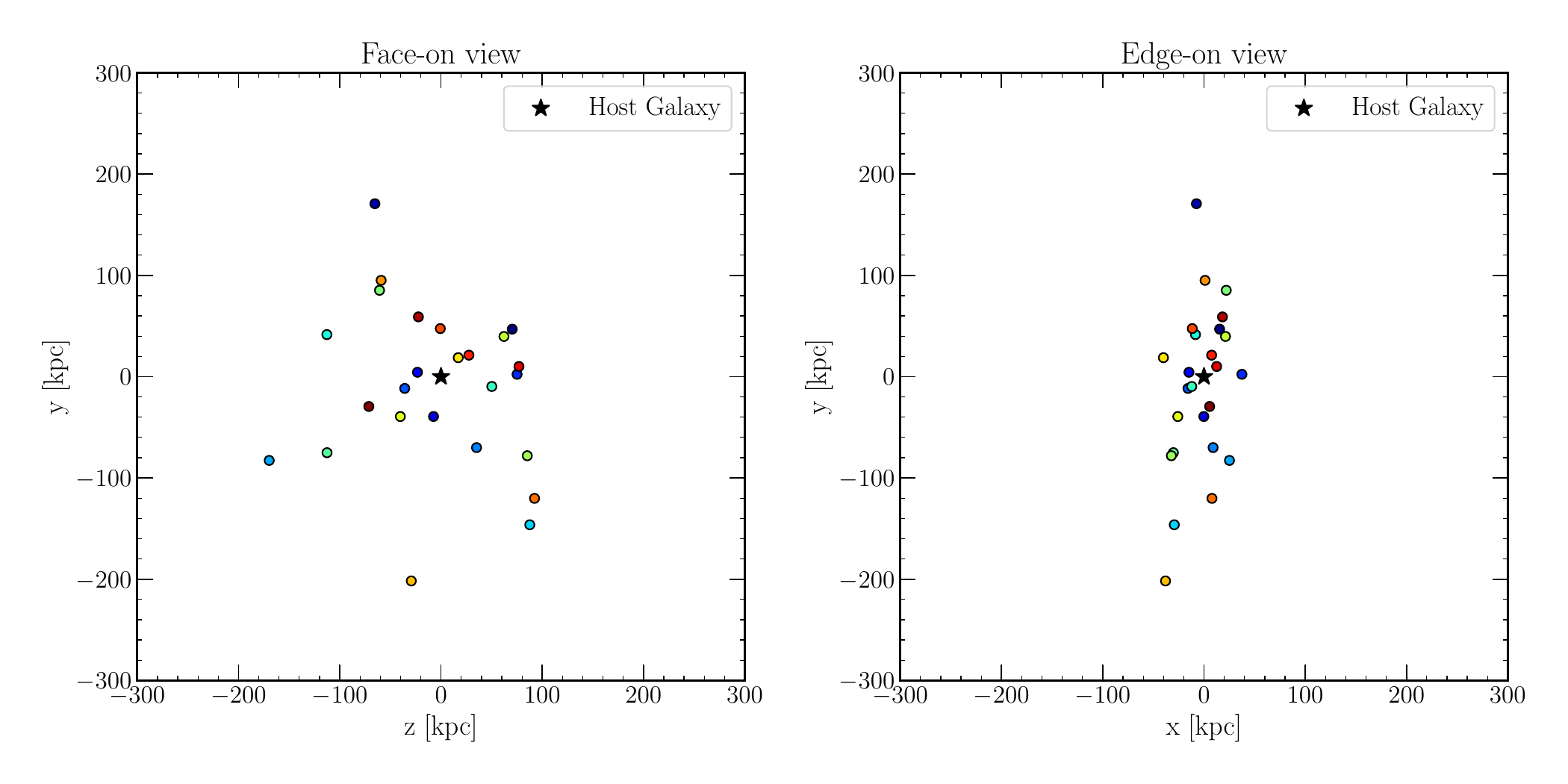}
     \caption{Face-on (left) and edge-on (right) views of $N_{\text{sat}}$ = 25 randomly generated test satellites, each represented by a different color. The star symbol denotes the host galaxy.}
     \label{fig:initial}
\end{figure*}

Today, the Lambda cold dark matter ($\Lambda$CDM) cosmological model is widely accepted among researchers due to its ability to explain a range of astrophysical phenomena such as Big Bang nucleosynthesis, the accelerated expansion of the Universe \citep{riess1998observational}, the power spectrum of the cosmic microwave background \citep{spergel2003first}, large-scale structure \citep{peacock2001measurement, tegmark2004three}, and the matter-energy budget of our Universe \citep{ade2016planck}. However, on smaller scales, it presents several challenges.  Among these, the plane of satellite galaxies has consistently remained a topic of debate \citep{pawlowski2018planes}, and such structures are considered highly improbable within the standard $\Lambda$CDM cosmological model \citep{kroupa2005great, pawlowski2021phase}. Several solutions have been proposed to explain this discrepancy; however, no general consensus has been reached.  These proposed solutions include accretion along filaments of the cosmic web \citep{libeskind2011preferred, lovell2011link}, infall of satellite galaxies in groups
\citep{li2008infall, d2008small, metz2009did, shao2018multiplicity, wang2013spatial, julio2024}, galaxy mergers \citep{smith2016,kanehisa2023b}, effects induced by the infall of a massive LMC \citep{garavitocamargo2021,pawlowski2022,vasiliev2023effect}, tidal dwarf galaxies \citep{pawlowski2011making, fouquet2012does, banik2022}, and hydrodynamics and baryonic physics \citep{pawlowski2015persistence,ahmed2017role, muller2018whirling, pawlowski2019halos}. Thus, while each hypothesis contributes valuable insight, a unified explanation remains elusive, highlighting the complexity of galaxy formation and evolution processes.

Precise investigation of the orbits of satellite galaxies requires a thorough understanding of several critical parameters. These parameters encompass precise measurements of a satellite's position, distance, line-of-sight velocity, proper motions, the underlying gravitational potential of the host galaxy, the Galactocentric distance of the Sun, and its motion relative to the Galactic center. However, many of these key parameters remain inadequately constrained, introducing a significant uncertainty in orbital analyses \citep{bland2016galaxy, joshi2007displacement, schonrich2010local}. Measurement uncertainties, especially in proper motions, can significantly affect analyses \citep{pawlowski2021s}. These uncertainties can be broadly classified as statistical or systematic \citep{li2021gaia}. Statistical uncertainties arising from random measurement variations can be mitigated through statistical methods and the use of larger datasets. Conversely, systematic uncertainties, which result from inherent biases in the measurement process or instruments,  pose more fundamental challenges.

\cite{maji2017there} demonstrate that the Milky Way’s disk of satellites (DoS) is dynamically transient. Their orbit modeling shows that the 11 classical satellites move away from the current DoS, causing it to thicken from \(c/a \sim 0.18\) (height \(\sim 19.6\) kpc) to \(c/a \sim 0.36\) (45 kpc) in 0.5 Gyr, and \(c/a \sim 0.42\) (64 kpc) in 1 Gyr. This evolution persists across the two tested galactic potentials: one replacing the stellar disk with a single-component Miyamoto–Nagai potential and another using only an Navarro–Frenk–White (NFW) dark matter halo potential. However, \cite{pawlowski2017considerations} notes that \cite{maji2017there} overlook measurement uncertainties, such as those related to proper motions, and argues that ignoring observational biases further undermines the reliability of their conclusions regarding the plane of satellite galaxies. Similarly, \cite{sawala2023milky} argues that the plane of satellites is more likely to be transient rather than rotationally supported, because backward and forward integration of satellite galaxy orbits results in a widening of the inferred plane thickness. However, they do not account for systematic errors in Gaia proper motions. These uncertainties are crucial because they can significantly affect the measured values and potentially alter the results. Without accounting for these uncertainties, the conclusions drawn about the stability and alignment of the Milky Way's satellite plane might be overly optimistic or pessimistic. This could lead to misinterpretations about the dynamical status of these satellite structures, thereby influencing our understanding of the formation and evolution of the Milky Way and its satellite system. Incorporating measurement uncertainties, particularly in proper motion data, is essential to achieve a more realistic and reliable understanding of the satellite galaxies' orbital behaviors. This is especially true when studying the joint behavior of an ensemble of satellite galaxies, since errors can conceal out intrinsic correlations (and approaches such as Monte Carlo sampling from the uncertainties effectively apply the errors twice; \citealt{pawlowski2021s}). 

To address this, we explore how different levels and types of measurement uncertainties affect the inferred stability of planes of satellite galaxies. Specifically, we employed computer simulations to model the orbital evolution of test satellite galaxies around a host galaxy that were set up as intrinsically stable satellite planes. We mock-observed these by applying controlled degrees of observational errors on their evolved positions and velocities, and then backward-integrated them to determine how well the intrinsically stable dynamical evolution can be inferred.

This paper is structured as follows. In Sect. \ref{Sect:Method}, we describe the methodology, including setting up an artificial satellite plane, the plane-fitting routine, the orbit integrations performed, the application of mock measurement errors, and the subsequent backward integration of the test satellites. Section \ref{Sect:Results} presents our results for various combinations of measurement errors, assumed Milky Way potentials, and initial orbital eccentricity distributions. We end with a discussion and conclusions in Sect. \ref{Sect:conclusion}.

\begin{table}[t]
\caption{Potential models}
\label{Table:Potentials} 
\centering
\begin{tabular}{c c c c }
\hline\hline
Potential Model & Halo Mass \\
\hline
    $ MW_{\text{low}}$  & $4.8 \times 10^{11} M_{\odot}$ \\
    $ MW_{\text{fiducial}}$ & $8 \times 10^{11} M_{\odot}$ \\
    $ MW_{\text{high}}$ & $11.2 \times 10^{11} M_{\odot}$ \\ 
\hline
\end{tabular}
\end{table}

\section{Methodology}
\subsection{Plane setup}

To study the evolution of artificial planes of satellite galaxies, the test satellites must be appropriately arranged around the host galaxy to initially form a planar structure. To achieve this, we used the Milky Way and its satellite distribution as a model to establish an artificial plane of satellite galaxies.  The Milky Way satellite plane, VPOS, spans approximately 250 kpc in radius, with a root mean square thickness ranging from 20 to 30 kpc. Therefore, the radial distribution, \( R \), of   \( N_{\text{sat}} \)  test satellites ranges from 20 to 250 kpc, while the vertical distribution extends from 0 to 20 kpc  \citep{pawlowski2018planes, pawlowski2021phase}. To generate a random radial distribution, we used the Von Neumann rejection algorithm, also referred to as rejection sampling \citep{neumann1951various}. This statistical technique facilitates the generation of random samples from a desired probability distribution. Figure \ref{fig:initial} shows the face-on and edge-on views of the randomly distributed \( N_{\text{sat}} = 25 \) test satellites around a host galaxy, initially set up as a planar structure. Here, the satellite plane is set perpendicular to the host galaxy's stellar disk. To produce the radial distance distribution, we used a power-law distribution function:
\begin{equation}
f(r) = r^{\alpha}
\label{funct}.
\end{equation}

Here, for $\alpha$, we used $-3$. For the velocities, we first defined the unit vectors for each test satellite in the radial, tangential, and perpendicular directions, with respect to the satellite plane. 
In vector representation, these are written as

\begin{align}
\mathbf{V_{\text{rad; i}}} & = |\mathbf{V_{\text{rad; i}}}| \cdot \hat{r}_{i} \\
\mathbf{V_{\text{tan; i}}} & = |\mathbf{V_{\text{tan; i}}}| \cdot \hat{t}_{i} \\
\mathbf{V_{\text{perp; i}}} & = |\mathbf{V_{\text{perp; i}}}| \cdot \hat{p}_{i}\,,
\end{align}

where $\mathbf{V_{\text{rad; i}} }$, $\mathbf{V_{\text{tan; i}} }$, and $\mathbf{V_{\text{perp; i}} }$ represent the radial, tangential, and perpendicular velocity vectors of the \(i\)-th test satellite, respectively. The quantities $|\mathbf{V_{\text{rad; i} }}|$, $|\mathbf{V_{\text{tan; i}} }|$, and $|\mathbf{V_{\text{perp; i} }}|$ denote the magnitudes of the radial, tangential, and perpendicular velocity, respectively. The unit vectors $\hat{r}_{i}$, $\hat{t}_{i}$, and $\hat{p}_{i}$ correspond to the radial, tangential, and perpendicular directions for the \(i\)-th test satellite.

\cite{hammer2021gaia} and \cite{cautun2017tangential} investigated the relations between radial and tangential velocities of Milky Way satellite galaxies and report a tangential bias. Specifically, \cite{cautun2017tangential} show that nine out of ten satellites with measured proper motions exhibit highly tangentially biased motions, with 80\% or more of their orbital kinetic energy attributed to tangential motion. Given the significance of tangential velocity, we initially set both $|\mathbf{V_{\text{perp; i}}}|$ and $|\mathbf{V_{\text{rad; i}}}|$ to zero, indicating no motion in the radial and perpendicular directions. Thus, only the tangential component is relevant. Satellites were initialized with a tangential speed equal to the circular velocity of the $MW_{\text{fiducial}}$ gravitational potential (see Table \ref{Table:Potentials}), guaranteeing circular orbits. To make the orbits eccentric, we rotated the tangential unit vectors of all satellites by an angle $\theta$ around the perpendicular unit vector. For each test satellite, the values for $\theta$ were drawn from a uniform distribution with a maximum range of $-\theta_\mathrm{max}$ and $+\theta_\mathrm{max}$, as listed in Table \ref{table:initial_parameters}. 

Once we obtained the radial $\mathbf{V_{\text{rad; i}}}$, perpendicular $\mathbf{V_{\text{perp; i}}}$, and tangential $\mathbf{V_{\text{tan; i}}}$ velocity components, we converted them into a Cartesian coordinate system. This allowed us to fully describe the motion of the test satellites in the simulated galaxy system.

\subsection{Plane fitting}

Once the test satellites were distributed around their host galaxy, the next step was to perform plane fitting to analyze their spatial arrangement. This technique allowed us to determine whether the satellites were arranged in a planar structure or scattered more isotropically around the host galaxy. For this purpose, we used an unweighted plane-fitting technique \citet{metz2007spatial}. This method begins with the calculation of the moment-of-inertia tensor of the satellite distribution, which is followed by its diagonalization. The first step of the process is to determine the centroid $r_{0}$, of the data points:

\begin{equation}
    r_{0} = \frac{1}{N} \sum_{i=1}^{N} r_{i}. 
\end{equation}

Subsequently, an eigenvalue analysis of the moment of inertia tensor, $\mathbf{T}_0$, is performed for the position vectors, \(\hat{r}_i = r_i - r_{0}, i = 1 \ldots n\) \citep{metz2007spatial, pawlowski2013dwarf}:

\begin{equation}
    T_0 = \sum_{i=1}^N \left[(\mathbf{r}_{i} - \mathbf{r}_{0})^2 \cdot \mathbf{1} - (\mathbf{r}_{i} - \mathbf{r}_{0}) \cdot (\mathbf{r}_{i} - \mathbf{r}_{0})^T\right].
\end{equation}

Here, \textbf{1} denotes the unit matrix and \(\mathbf{r}^{\textbf{T}}\) is the transposed version of the vector \(\mathbf{r}\). The square root of the eigenvalues of the moment of inertia indicates the extent along three axes \textit{(a, b, c)} of the fitted ellipsoid to the satellite distribution. These values are proportional to the rms deviation relative to the eigenvectors of \textbf{T}. The eigenvector corresponding to the largest eigenvalue defines the normal of the plane, encompassing the centroid. Meanwhile, the eigenvectors corresponding to the intermediate and smallest eigenvalues indicate the directions of the intermediate and major axes of the distribution, respectively. Consequently, we determined the axial ratios \textit{c/a} and \textit{b/a}.  A small value of \textit{c/a} suggests two possibilities: if \textit{b/a} is large, it indicates an oblate distribution resembling a thin plane; if \textit{b/a} is similarly small \textit{($c/a \approx b/a$)}, it suggests a narrow, prolate distribution resembling a filament-like shape \citep{metz2007spatial}.

 \begin{figure*}[t]
\centering
   \includegraphics[width=18cm]{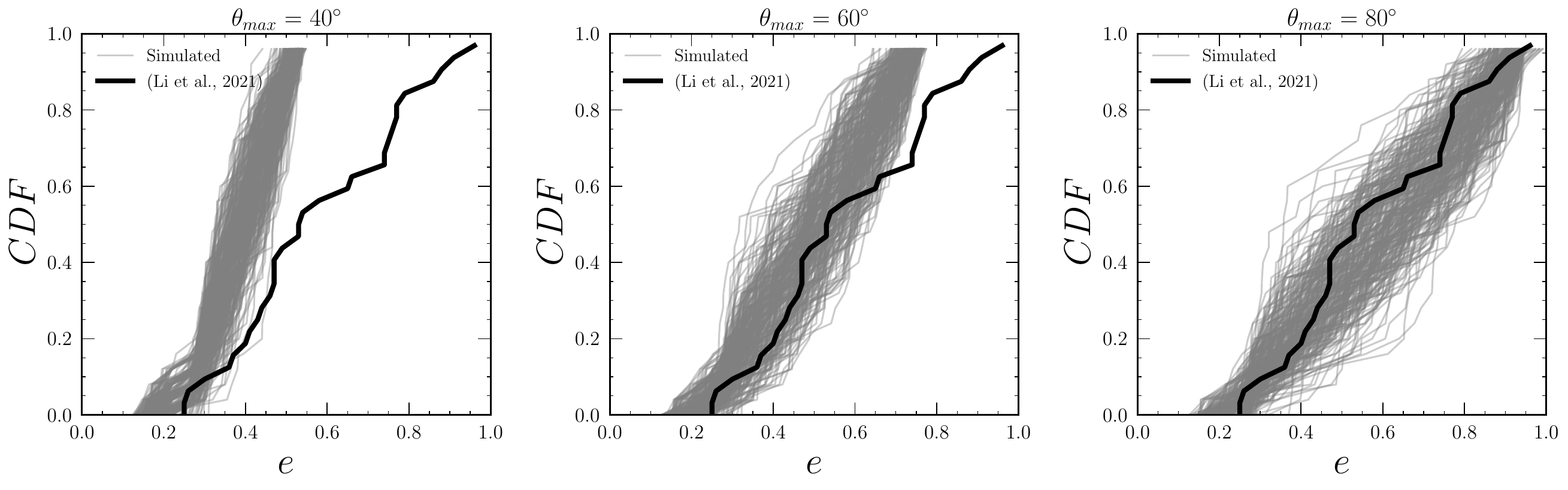}
     \caption{Comparison of eccentricity $(e)$ distributions across different $\theta$. The figure presents the eccentricity CDF for three different $\theta_{max}$: 40\textdegree, 60\textdegree, and 80\textdegree, from left to right, respectively, shown in gray. The solid black line in each panel represents the observed data for Milky Way satellites from \citep{li2021gaia}, while the overlaid gray CDFs correspond to simulated results for $N_{\text{sat}}$ test satellites.}
     \label{fig:e}
\end{figure*}

\subsection{Adopted potential}

Understanding the gravitational potential and mass distribution of the Milky Way is crucial to numerically integrating the orbits of test satellites. Despite extensive research over the past century, the exact distribution of the Milky Way's mass remains a topic of debate. Current estimates of the total mass of the Milky Way range between $0.5 - 3 \times 10^{12} M_{\odot}$ \citep{wang2020mass}, varying based on the methodology and the underlying assumptions \citep{bland2016galaxy}. Therefore, significant uncertainty surrounds the mass of the Milky Way, necessitating consideration of these ambiguities in mass distribution. 

For this study, we used one fiducial potential for forward integration and three distinct gravitational potentials, as detailed in Table \ref{Table:Potentials}, to investigate their effects on the evolution and inferred stability of the plane of satellite galaxies in backward integration. This allowed us to evaluate the effect of adopting an incorrect Milky Way mass on the inferred satellite plane's evolution. To numerically integrate orbits, we relied on \texttt{Galpy} \citep{bovy2015galpy} to investigate the orbital properties of the test satellites. Within the \texttt{galpy.potential} module, the \texttt{MWPotential2014} model represents the Milky Way's gravitational potential and is obtained by fitting to a wide range of observational data. This simplified static potential model comprises the following components:

A spherically symmetric power-law density potential with an exponential cut-off for the bulge:
\begin{equation}
\rho(r) = \rho_{0} \, r^{-\alpha} \exp \left( -\left( \frac{r}{r_{c}} \right)^{2} \right),
\end{equation}
with a power-law index of \(\alpha = 1.8\) and a cut-off radius of \(r_{c} = 1.9\) kpc. The disk component is modeled using the axisymmetric Miyamoto–Nagai disk \citep{miyamoto1975three}:
\begin{equation}
\Phi(R,z) = -\frac{\Phi_{0}}{\sqrt{R^2+(a+\sqrt{z^2+b^2})^2}},
\end{equation}
where \(a = 3\) kpc and \(b = 0.28\) kpc.

The dark matter halo is characterized by the NFW halo \citep{navarro1996structure}:
\begin{equation}
\rho(r) = \frac{\rho_{0}}{(r/h)(1+r/h)^2},
\end{equation}
with a characteristic radius \(h = 16\) kpc.

The \texttt{MWPotential2014} model is scaled such that the circular velocity at a distance of 8 kpc from the galactic center within the disc (\(z = 0\) kpc) is 220 km/s, with a virial mass of approximately \(8 \times 10^{11} M_{\odot}\). In this study, we refer to this potential as the fiducial potential, \(MW_{\text{fiducial}}\). For the two remaining Milky Way potentials, we adjusted \(MW_{\text{fiducial}}\) by rescaling the halo mass to achieve the desired mass distribution. Specifically, the \(MW_{\text{low}}\) model was configured with a halo mass 40\% lower than that of \(MW_{\text{fiducial}}\), while \(MW_{\text{high}}\) had a halo mass 40\% higher than that of \(MW_{\text{fiducial}}\), thereby providing deeper insights into the influence of mismatching the Milky Way's mass on the inferred behavior and stability of satellite galaxy planes.

\begin{table}[h]
\caption{Parameters for forward integration models}
\label{table:initial_parameters} 
\centering
\begin{tabular}{ccccccc }
\hline\hline

Model & $N_{\text{sat}}$ & $\theta_{max}$  & Potential Model & $N_{\text{realization}}$ \\
\hline
FI-01 & 25 & 40\textdegree  & $ MW_{\text{fiducial}}$ & 20 \\
FI-02 & 25 & 60\textdegree  & $ MW_{\text{fiducial}}$ & 20 \\
FI-03 & 25 & 80\textdegree  & $ MW_{\text{fiducial}}$ & 20 \\

\hline
\end{tabular}
\tablefoot{ The table lists the model numbers (FI), the number of satellites (\(N_{\text{sat}}\)), the maximum angle \(\theta_{max}\) in degrees, the gravitational potential model (\(\textit{MW}_{fiducial}\)), and the number of realizations (\(N_{\text{realization}}\)) for each forward-integration scenario.}
\end{table}

\subsection{Forward Integration}
We conducted forward integration of the satellite system over a 5 Gyr time interval for $N$ = 20 random realizations of the initial setup. We selected three distinct maximum tangential velocity angles, $\theta_{\text{max}}$ = 40\textdegree, 60\textdegree, and 80\textdegree, for forward integration, as shown in Table \ref{table:initial_parameters}. Notably, larger angles allow for more eccentric orbits. 

The resulting cumulative distribution functions (CDFs) of eccentricity for the three different values of $\theta$ are shown in Fig. \ref{fig:e}. We also overlaid, as a solid black curve, the eccentricity data for observed Milky Way satellites from \citep{li2021gaia} onto the plot, which contains data for 46 satellite galaxies from \cite{fritz2018gaia}, \cite{mcconnachie2020updated} and \cite{simon2019faintest}. However, for our comparison, we selected 35 satellite galaxies, as 11 had eccentricities $e \gg 1$. This addition allowed us to compare the simulated eccentricities with those of observed Milky Way satellite galaxies. From the comparison, it is apparent that $\theta_{\text{max}}$ = 80\textdegree aligns closely with the observed eccentricities of satellite galaxies. Each forward integration was conducted using the $ MW_{\text{fiducial}}$ potential model. Figure \ref{fig:e} also shows that the CDF range for \cite{li2021gaia} spans approximately 0.2 to 1.0. For the simulated eccentricities of our initial test runs, the range was between 0.0 and 1.0, because circular orbits were also included. To align the simulated eccentricities with the observed data, we excluded values ranging between $\theta$ = -20\textdegree and 20\textdegree to ensure that perfectly circular orbits were not generated.

\subsection{Uncertainties}

\begin{figure*}[t]
\centering
   \includegraphics[width=17cm]{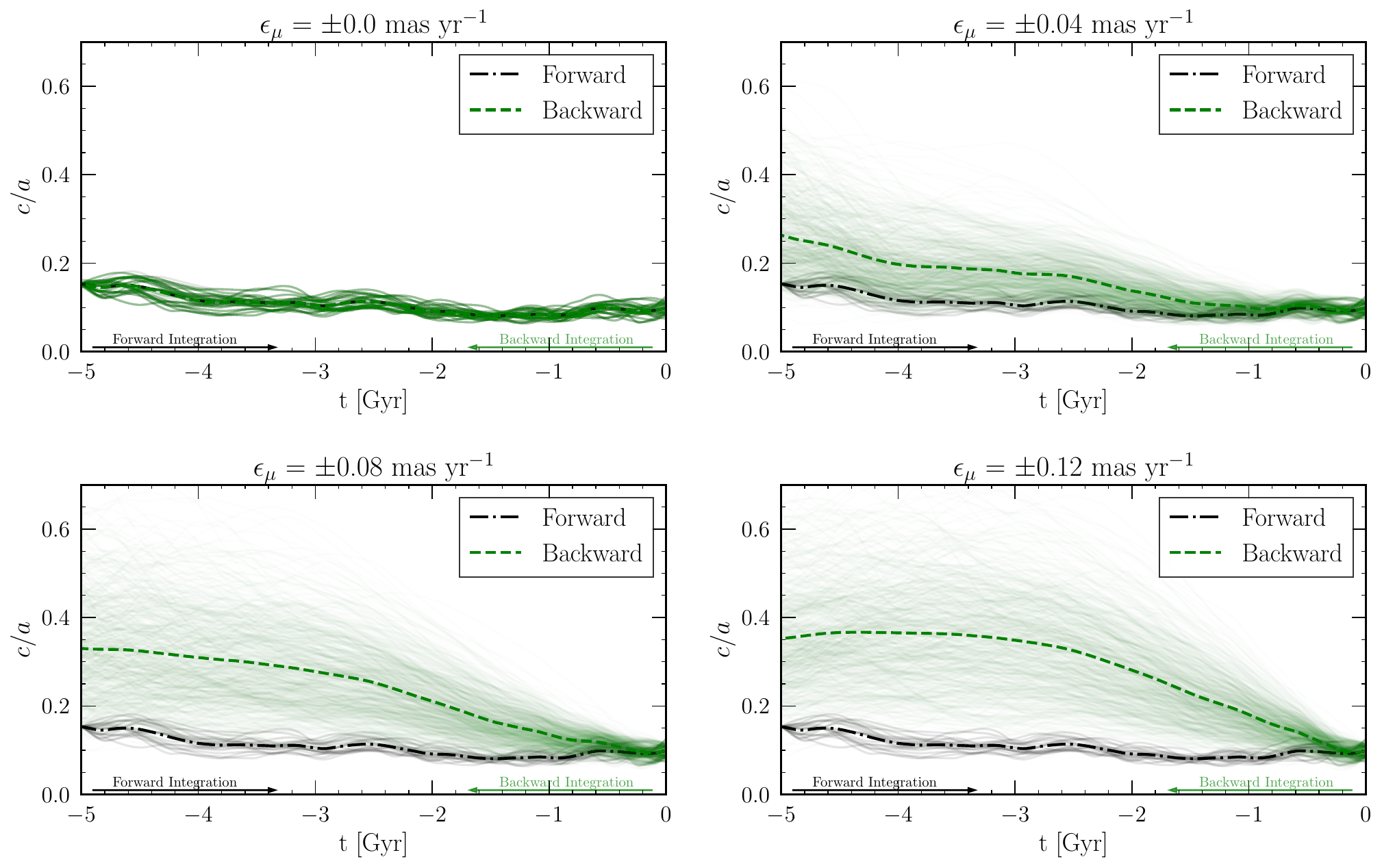}
   \caption{Effect of proper motion uncertainties on the plane of satellite galaxies. Each panel shows $N_{\text{realization}}$ forward integrations as black curves with the mean indicated by a dot-dashed line, and $M_{\text{realization}}$ backward integrations as green curves with the mean indicated by a dashed line.}     
\label{fig:only_pm}
\end{figure*}

To precisely determine the motion of a satellite galaxy, its position, distance, and 3D motion are crucial. Spectroscopy is used to track the motion of a galaxy via the Doppler effect. However, it is essential to note that radial velocity measures only one dimension. To determine how a galaxy moves tangentially across the sky, proper motions are measured. The Hubble Space Telescope (HST) has measured the proper motion of most of the 11 brightest classical satellite galaxies of the Milky Way \citep{redd2018astronomers}. Since then, largely due to the advent of Gaia, the availability of proper motion measurements has improved significantly \citep{gaia2016gaia, brown2016gaia}. However, highly accurate proper motion measurements remain limited and are available only for the more massive, luminous, or nearby satellite galaxies. 

According to Gaia Data Release 2 (DR2), the random proper motion errors for Milky Way satellites range from approximately 0.006 to 0.641 mas yr\(^{-1}\) \citep{fritz2018gaia}, and the corresponding systematic errors are estimated to be approximately 0.035 to 0.066 mas yr\(^{-1}\) \citep{gaia2018gaia, helmi2018gaia}. In Gaia Early Data Release 3 (EDR3), the uncertainties in proper motions, derived from both statistical and systematic errors, have approximately halved\citep{Lindegren2021,mcconnachie2020updated,li2021gaia,Battaglia2022}.

In this study, we investigated four different ranges of proper motion uncertainties: $\epsilon_\mu = 0.00$\,mas yr$^{-1}$, 0.04 mas yr$^{-1}$, 0.08 mas yr$^{-1}$, and 0.12 mas yr$^{-1}$. We specifically opted for 0.00 mas yr$^{-1}$ proper motion uncertainties to represent cases where proper motion was measured with the highest accuracy. This condition allowed us to examine the evolution of the plane of satellite galaxies under various parameters, independent of proper motion errors.  We randomly sampled these proper motion uncertainties from Gaussian distributions with width $\epsilon_\mu$.

For distance uncertainties, we chose two specific values: $\epsilon_\mathrm{dist} = 0$\% and 5\%. Distances are crucial factors in understanding the spatial distribution and dynamics of satellite galaxies around the Milky Way, as well as their effects on the inferred stability of the plane of satellite galaxies. Again, distance errors were sampled from a Gaussian with width $\epsilon_\mathrm{dist}$.

\subsection{Backward integration}

After completing the forward integration, we recorded the orbital parameters, such as right ascension, declination, proper motion in right ascension and declination, distance, and line-of-sight velocity, for each test satellite system at the final snapshot of all simulation realizations. We then mocked-observed these systems by adding uncertainties in proper motions and distance such that

\begin{align}
\mu_{\alpha*_{\text{new}}} & = \mu_{\alpha*} + \epsilon_{\mu} \label{eq:mu_alpha_new} \\
\mu_{\delta_{\text{new}}} & = \mu_{\delta} + \epsilon_{\mu}  \label{eq:mu_delta_new} \\
\text{dist}_{\text{new}} & = \text{dist} + \epsilon_{\text{dist}} \label{eq:d_new}.
\end{align}

Equations \eqref{eq:mu_alpha_new} -- \eqref{eq:d_new} provide the updated proper motions and distance parameters for each test satellite, incorporating their respective uncertainties. The values of $\epsilon_{\mu}$ were drawn from a Gaussian distribution. In contrast, for $\epsilon_{\text{dist}}$, a percentage error was drawn from a Gaussian distribution, so that more distant satellites had larger absolute distance uncertainties. Table \ref{table:4} details the ranges of uncertainties in these parameters. After performing mock observations, we integrated the system backward for 5 Gyr. We repeated this process 30 times for each of the 20 forward integrations, using different randomly generated errors each time, resulting in a total of 600 backward integrations to return the test satellite system to its initial state. We conducted this procedure using all three Milky Way potential models. Each row in Table \ref{table:4} represents a combination of various parameters that could influence the inferred stability of the satellite plane. Overall, with four proper motion uncertainties, two distance uncertainties, and three potential models, there were 24 model combinations for each $\theta_{max}$. Each model underwent 600 backward integrations, totaling 43,200 integrations in total.

\section{Results}

 In this section we present the main results of the simulation \footnote{GitHub repository of the code used to generate the results: \href{https://github.com/astro-pkumar/PlaneOfSatelliteGalaxies/tree/main}{Plane of Satellite Galaxies}}. For brevity, and because it most closely resembles the eccentricities of the observed Milky Way satellite galaxies, we show only results for $\theta_{max} = 80$ \textdegree. The results for the other $\theta_{\text{max}}$ are qualitatively similar and are provided in Appendix \ref{Appendix}.

\label{Sect:Results}

\subsection{Effect of proper motion uncertainties}

Figure \ref{fig:only_pm} presents results that focus on introducing proper artificial motion and distance uncertainties to the test satellite system before backward integration. These results are divided into four panels, each corresponding to a distinct level of proper motion uncertainty.

Each panel shows an array of curves, in gray and green, indicating the \(N_{\text{realization}}\) and \(M_{\text{realization}}\) numbers for forward and backward integration, respectively. The average value at each time step is shown as a dashed-dotted curve, with arrows pointing to the right and left. In this case, we adopt \(MW_{\text{fiducial}}\) for both forward and backward integrations, without introducing any distance uncertainties.
In the upper left panel, when $ \epsilon_{\mu} $ = 0.00 mas yr$^{-1}$, the mean minor-to-major ratio, $c/a$, obtained from backward integration, aligns perfectly well with the mean minor-to-major ratio obtained from forward integration, as each forward integration curve is overplotted by backward integration. This demonstrates a high degree of predictability and confirms that the orbital plane remains stable and unchanged in scenarios where the proper motion is measured with perfect accuracy. It also implies that, at least in the theoretical framework of the simulation, the inherent dynamics of the satellite system are sufficiently deterministic to allow precise backtracking of its orbital path, provided that the initial conditions are precisely known. These results serve as a control scenario against which the other panels, which include proper motion errors, can be compared. Model BI-01 from Table \ref{table:4} shows similar behavior: even after 3 Gyr of backward integration, the difference between forward and backward, $\Delta_{c/a} = 0.0$, and the fractional change, $f_{c/a}$,  is 1.00, meaning that both forward and backward integration yield the same output when proper motion uncertainties are zero.

In the upper right panel, we have applied Gaia-level systematic uncertainties $ \epsilon_{\mu} $ =  0.04 mas yr$^{-1}$. Here, a slight uncertainty in proper motion leads to a visible divergence between the mean axial ratios of forward and backward integrations. The mean axial ratio from the backward integration does not align as closely with the forward integration as in the first case, when there were no uncertainties. This suggests a decrease in the predictability and inferred stability of the orbital plane with the introduction of proper motion errors. However, the curve appears to retain its shape to some degree. The effect of uncertainties is also evident from Model BI-02 in Table \ref{table:4}, which shows that with the introduction of $\epsilon_{\mu} = 0.04$ mas yr$^{-1}$, $\Delta_{c/a}$ shifts from 0.00 to 0.073, and $f_{c/a}$ becomes 1.69. Thus, the inferred past plane width is typically 70\% higher than the true value.

Increasing the proper motion uncertainties to 0.08 mas yr$^{-1}$ further increases the divergence between the means of backward and forward integration, as shown in the lower left panel. This indicates a further decline in the predictability and stability of the orbital plane as proper motion uncertainty increases. Moreover, Table \ref{table:4} also shows that in Model BI-03,  $\Delta_{c/a}$ shifts yet again, from 0.07 to 0.17, and $f_{c/a}$ rises to 2.64, meaning that the typical inferred plane width is almost three times larger than the true one.

Under extreme proper motion uncertainties, 0.12 mas yr$^{-1}$, the divergence between the mean forward and mean backward integration, shown by dashed curves, is more pronounced. The mean axial ratio from backward integration shows a significant divergence from the forward integration mean, implying a severe reduction in the orbital plane's predictability and stability with the highest proper motion uncertainty examined. 

Moreover, the original shape of the curve is no longer discernible and deviates significantly more from the forward integration curve compared to the other uncertainties examined. This suggests that the actual past evolution of the system cannot be reliably traced under such high uncertainties. This shows that the inferred dynamics of the system are highly sensitive to proper motion uncertainties, making it challenging to draw reliable conclusions about the satellites' orbital evolution under these conditions. This case is represented by Model BI-04 in Table \ref{table:4}, which shows that $\Delta_{c/a}$ = 0.24 and $f_{c/a}$ = 3.32. 

\subsection{Effect of intrinsic plane height}

\begin{figure*}[t]
\centering
   \includegraphics[width=17cm]{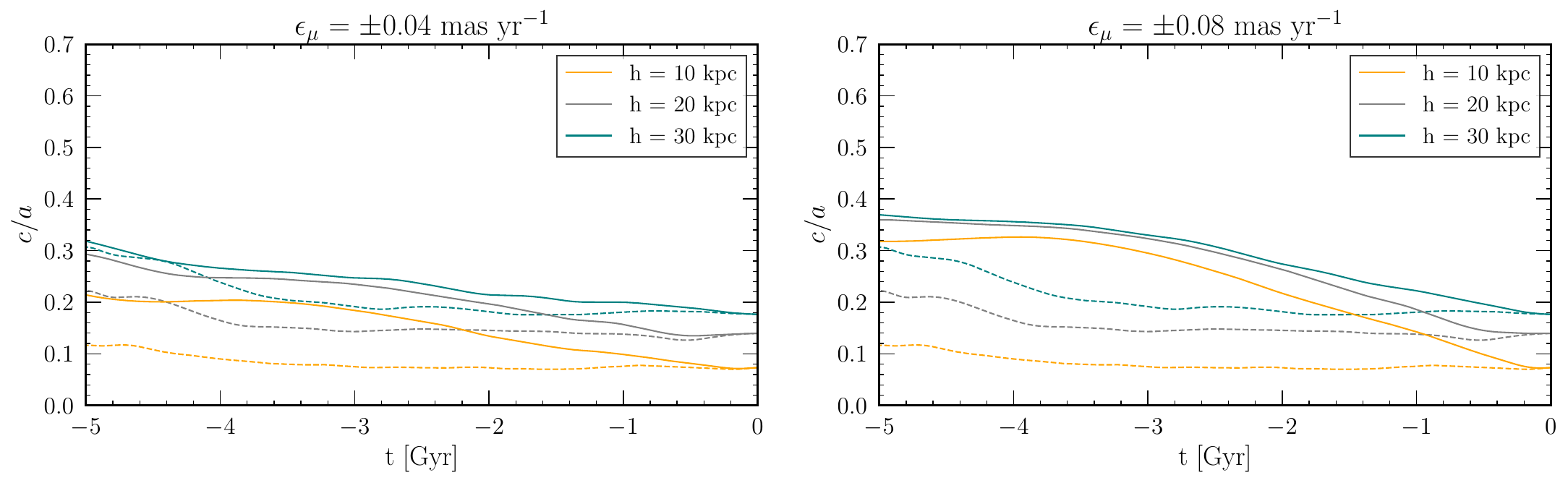}
\caption{Effect of proper motion uncertainties on satellite galaxies under different initial vertical distributions of $h = $ 10, $h$ = 20 kpc, and $h$= 30 kpc. The left panel shows proper motion uncertainties of 0.04 mas yr$^{-1}$, while the right panel displays uncertainties of 0.08 mas yr$^{-1}$. Dashed curves indicate forward integrations and solid curves represent backward integrations.}
\label{fig:with_diff_h}
\end{figure*}

We also tested whether the effect of proper motion uncertainties on the plane of satellite galaxies was sensitive to our choice of the initial vertical distribution of the intrinsic satellite planes. Specifically, we ran the system with three different plane heights — $h = 10 \, \text{kpc}$, $h = 20 \, \text{kpc}$, and $h = 30 \, \text{kpc}$ — bracketing the range of observationally inferred satellite plane heights for the Milky Way and M31 \citep{ibata2013vast, pawlowski2013dwarf}.

Figure \ref{fig:with_diff_h} displays two panels showing the results for adopted proper motion uncertainties of 0.04 mas yr$^{-1}$ and 0.08 mas yr$^{-1}$. In both plots, the dash-dotted curves represent mean axis ratios of the forward integrations for different heights of \( h = 10, 20, \) and \( 30 \) kpc, indicated by orange, gray and cyan, respectively. As expected, when the vertical distribution is lower, the mean axis ratio of the system is lower, making the system flatter compared to the other values of \( h \). In backward integration, after adding proper motion uncertainties, we observe that all three curves exhibit a similar trend: errors in proper motion lead to an apparent widening of the satellite plane. This effect is substantial in all three cases, indicating that our findings are insensitive to the exact choice of intrinsic plane height.

Table \ref{table:different_h} summarizes the quantitative impact of proper motion uncertainties on the inferred change in the mean axis ratio. For a lower intrinsic plane height of $h = 10\,$ kpc (model BI-L10 and BI-H10 in Table \ref{table:different_h}), the relative effect is stronger, with $f_{c/a} =$ 2.45 (3.94) compared to $f_{c/a} =$ 1.64 (2.25) for the model with $h =$ 20 kpc, for adopted proper motion uncertainties of 0.04 (0.08) mas yr$^{-1}$. Similarly, for an intrinsically wider plane (models BI-L30 and BI-H30 in Table \ref{table:different_h}), the relative change in plane flattening is slightly lower, with $f_{c/a} =$ 1.29 and 1.72 for proper motion uncertainties of 0.04 and 0.08 mas yr$^{-1}$, respectively. However, the absolute change in axis ratio, measured via $\Delta_{c/a}$, does not differ as significantly.

\begin{table}[b]
\caption{Parameters and results for backward integration models at 3 Gyr for different intrinsic plane heights, similar to Table \ref{table:4}}
\label{table:different_h} 
\centering
\begin{tabular}{ccccccc}
\hline\hline
Model & $\epsilon_{\mu}$ [mas yr$^{-1}$] & $h$ [kpc] & $\Delta_{c/a}$ & $f_{c/a}$  \\
\hline

BI-L10 &            $\pm$ 0.04  &     10     & 0.10 $\pm$ 0.06  & 2.45 $\pm$ 0.84  \\
BI-L20 &            $\pm$ 0.04  &     20     & 0.09 $\pm$ 0.07   & 1.64 $\pm$ 0.51  \\
BI-L30 &            $\pm$ 0.04  &     30     & 0.06   $\pm$ 0.06   & 1.29 $\pm$ 0.32  \\
\hline
BI-H10 &            $\pm$ 0.08  &     10     &  0.19 $\pm$ 0.09  & 3.94 $\pm$ 1.32 \\
BI-H20 &            $\pm$ 0.08  &     20     & 0.18 $\pm$ 0.09   & 2.25 $\pm$ 0.75  \\
BI-H30 &            $\pm$ 0.08  &     30     & 0.14 $\pm$ 0.09  & 1.72 $\pm$ 0.49   \\

\hline
\end{tabular}
\end{table}

\subsection{Effect of distance uncertainties}
\label{subsec_dist_err}

\begin{figure*}[t]
\centering
   \includegraphics[width=17cm]{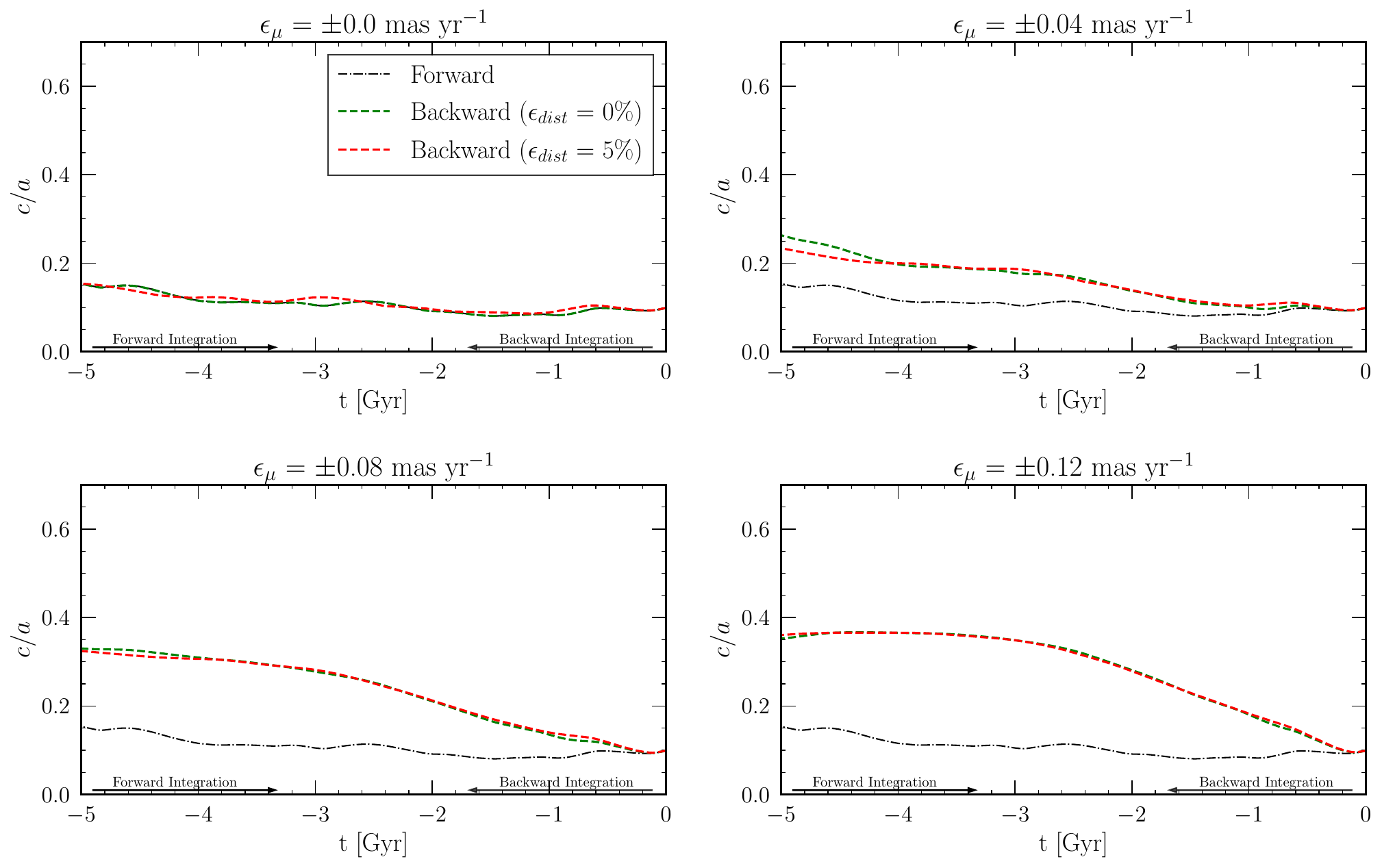}
\caption{Similar to Fig. \ref{fig:only_pm}, but including a comparison with 5\% distance uncertainties in the system. The green curve in each panel represents 0\% distance uncertainties, while the red curve shows 5\%. Proper motion uncertainties are: 0.00 mas yr$^{-1}$ (upper left), 0.04 mas yr$^{-1}$ (upper right), 0.08 mas yr$^{-1}$ (lower left), and 0.12 mas yr$^{-1}$ (lower right). Both forward and backward integrations are performed using the $MW_{\text{fiducial}}$ potential model.}
\label{fig:dist_pm}
\end{figure*}

To further assess the impact of observational uncertainties, we introduced additional 5\% distance uncertainties prior to backward integration. This allowed us to study the behavior of the plane under both proper motion and distance uncertainties. In all four panels of Fig. \ref{fig:dist_pm}, backward integration includes two curves: one without distance uncertainties (shown in green) and one with 5\% distance uncertainties (shown in red). As shown in the upper left panel of Fig.  \ref{fig:dist_pm}, when proper motion uncertainties are ignored, distance uncertainties have a noticeable effect: the system does not return to its original state but instead experiences a slight change. Despite this, the mean axis ratios from the backward and forward integrations remain in good agreement, suggesting that the inferred fundamental orbital evolution is not significantly affected, even with the introduction of 5\% uncertainties in the distances of the test satellites. This is also evident in Table \ref{table:4}, where in models BI-01 and BI-13, both $\Delta_{c/a}$ and $f_{c/a}$ change only slightly after 3 Gyr of backward integration, from 0.00 to 0.02 and from 1.00 to 1.17, respectively. Thus, a 5\% error in the distances translates to a widening of the inferred past plane height over the true value, reaching up to 17\% at specific times (e.g., after 3 Gyr), although the deviation remains negligible for the majority of the 5 Gyr integration period.

As we introduce proper motion uncertainties, we observe that the relative effect of distance uncertainties fades away as the proper motion error becomes dominant. This is evident in the lower right panel of Fig. \ref{fig:dist_pm}, where the highest proper motion uncertainties result in both curves showing similar behavior. Moreover, although not included in this paper, we also performed a test for 10$\%$ distance uncertainties. We find that the effect on the system becomes more pronounced with increasing uncertainties. However, this effect diminishes as the uncertainties in proper motions increase.

We also performed additional tests to examine the correlation between proper motion errors and distance errors. We find that, although there is a noticeable change, the overall effect is not extreme. This is discussed in Appendix \ref{AppendixPart2_recomputing_corr}.

\subsection{Effect of adopted potential}
\begin{figure*}[b]
\centering
   \includegraphics[width=17cm]{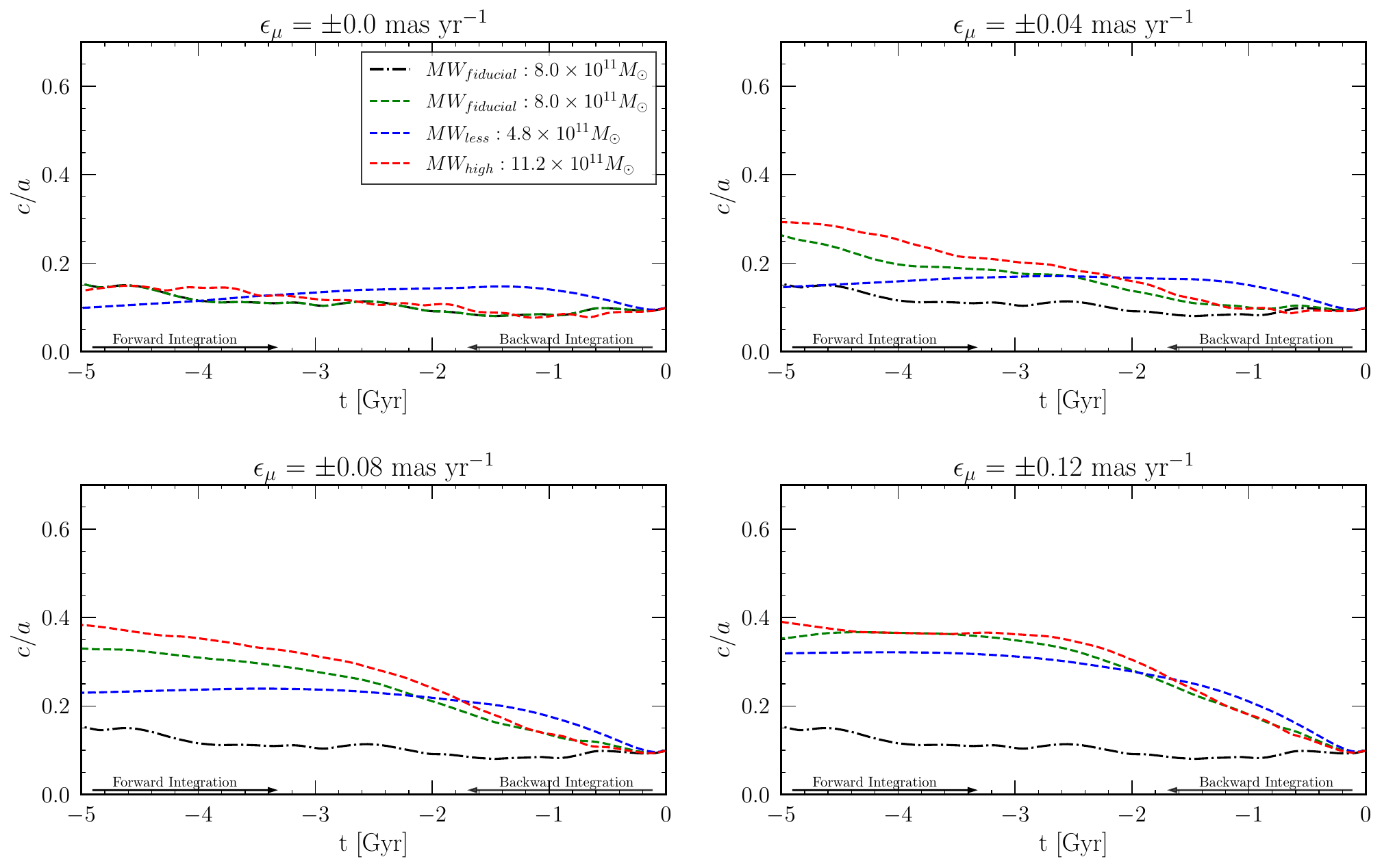}
     \caption{Evolution of the plane of satellite galaxies with proper motion uncertainties included in each panel. The black dashed-dotted curves represent the mean axis ratio for forward integration, while the colored dashed curves show the mean axis ratio for backward integration under different potential models characterized by halo mass. Specifically, the blue, green, and red dashed curves correspond to the ${c/a}$ values under the $MW_{\text{low}}$, $MW_{\text{fiducial}}$, and $MW_{\text{high}}$ potential models, respectively. Note that these results do not account for distance uncertainties.}
     \label{fig:80_all_mass_all_pmra_no}
\end{figure*}

\begin{figure*}[t]
\centering
   \includegraphics[width=17cm]{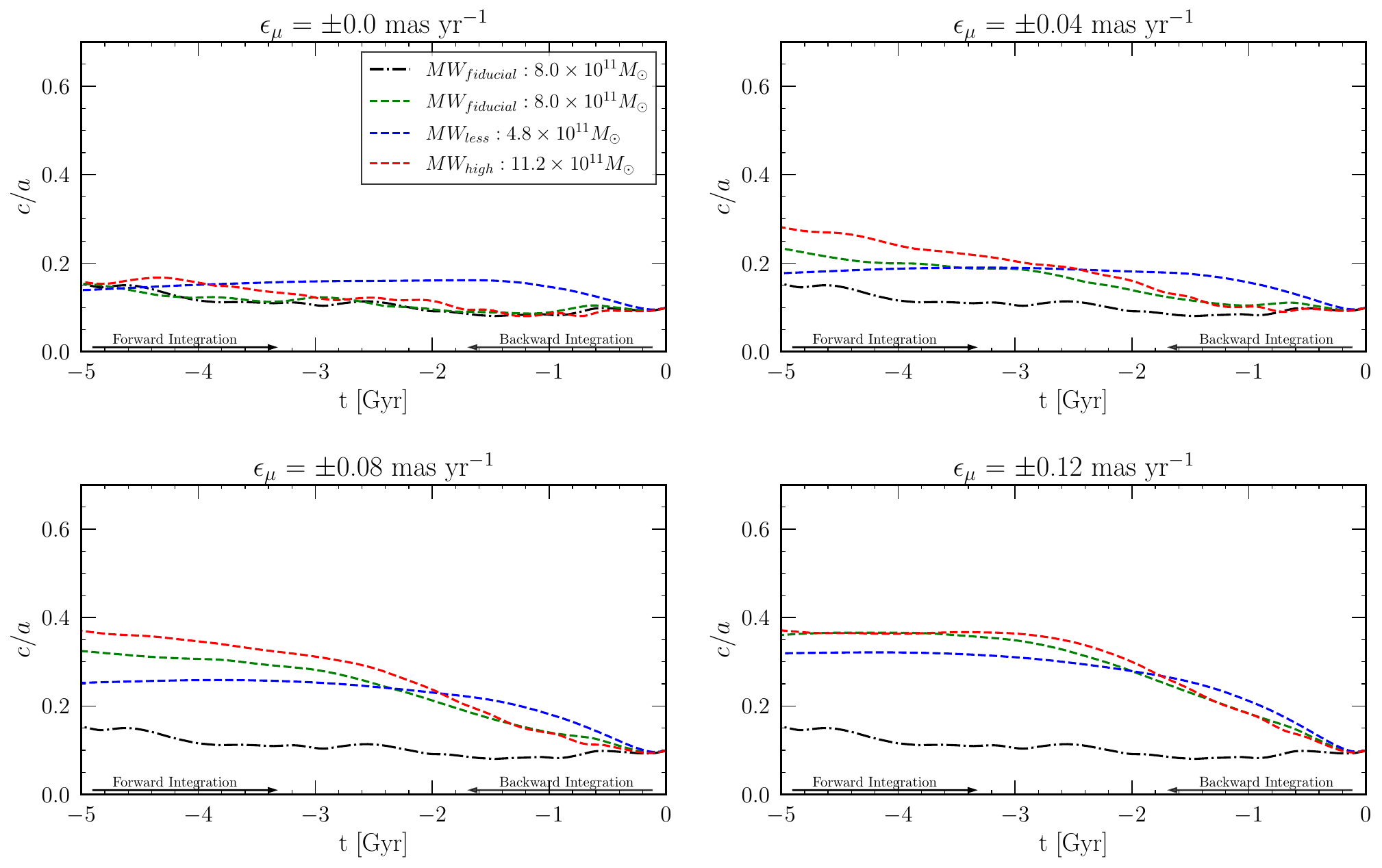}
     \caption{Similar to \ref{fig:80_all_mass_all_pmra_no}, but here we include a $5\%$ uncertainty in distances.}
     \label{fig:80_all_mass_all_pmra_yes}
\end{figure*}

In the upper left panel of Fig. \ref{fig:80_all_mass_all_pmra_no}, without the uncertainties in proper motion and distance, the results provide a clear picture of the fundamental dynamical behavior of the satellite plane when subjected to different potential models. The colored curves depict the mean backward integration under different Milky Way potentials. The backward integration, particularly the green curve representing the $MW_{\text{fiducial}}$ potential, demonstrates perfect consistency with its forward counterpart, implying a stable evolution when the potential model is the same. This case has been discussed in the previous sections.

The red curve, representing the $MW_{\text{high}}$ model, exhibits a reasonable degree of stability, although it does not completely resemble the forward integration, suggesting some sensitivity to the host galaxy's mass assumptions. This difference is shown in BI-09 of Table \ref{table:4}, which shows $\Delta_{c/a}$ = 0.01 after 3 Gyr, and $f_{c/a}$ = 1.13. In contrast, the blue curve, representing the $MW_{\text{low}}$ mass model, shows a markedly different behavior. Its axis ratio initially increases before flattening, which could imply a dynamical history distinct from the correct one for the plane of satellites within such a lower mass potential model. Here, for the lower mass Milky Way potential, $\Delta_{c/a}$ = 0.03 after 3 Gyr, and $f_{c/a}$ = 1.27.

The other three panels of Fig. \ref{fig:80_all_mass_all_pmra_no} show the effect of both potential models and proper motion uncertainties on the stability of the plane of satellites. In the upper right panel, we have proper motion uncertainties of $\pm$0.04 mas yr$^{-1}$. The effect under $ MW_{\text{fiducial}}$ was already discussed in the previous section. For the $MW_{\text{high}}$ model, the trend is similar to $MW_{\text{fiducial}}$, but with slightly larger axis ratio values. The $MW_{\text{low}}$ model again shows a distinct behavior, where it initially rises and then flattens out over time.

When adding more uncertainties, up to Gaia levels and beyond, the effect strengthens in all the remaining panels, and the plane of the system becomes even wider. Under 0.12 mas yr$^{-1}$ proper motion uncertainty, all three systems show a similar trend, suggesting that higher proper motion uncertainties overshadow the effect of using an incorrect Milky Way model. This observation indicates that, at higher levels of proper motion uncertainty, the distinctions between the various Milky Way potential models become less pronounced, as the increased uncertainty tends to dominate the inferred dynamics of the satellite plane.

Building upon this, we introduced an additional layer of complexity by incorporating distance uncertainties. A comparison between Fig. \ref{fig:80_all_mass_all_pmra_no} and Fig. \ref{fig:80_all_mass_all_pmra_yes}, reveals a noticeable effect of distance uncertainties across all three potential models, but the evolution of the plane does not change drastically. However, in the presence of higher proper motion uncertainties, the effect of distance uncertainties is again overshadowed under all three potential models.  Models BI-13 to BI-24 show this behavior when 5\% uncertainties are induced.

We also verified that our results remain qualitatively unchanged when using different Milky Way gravitational potentials. Specifically, we performed the analysis with \texttt{MilkyWayPotential2022} from \texttt{Gala} \citep{price2017gala}, including variations in halo mass and proper motion uncertainties, finding no substantial differences in the overall trends (see Appendix \ref{AppendixPart2_recomputing_gala}).

\subsection{Analysis of individual axes}

\begin{figure*}[t]
\centering
   \includegraphics[width=17cm]{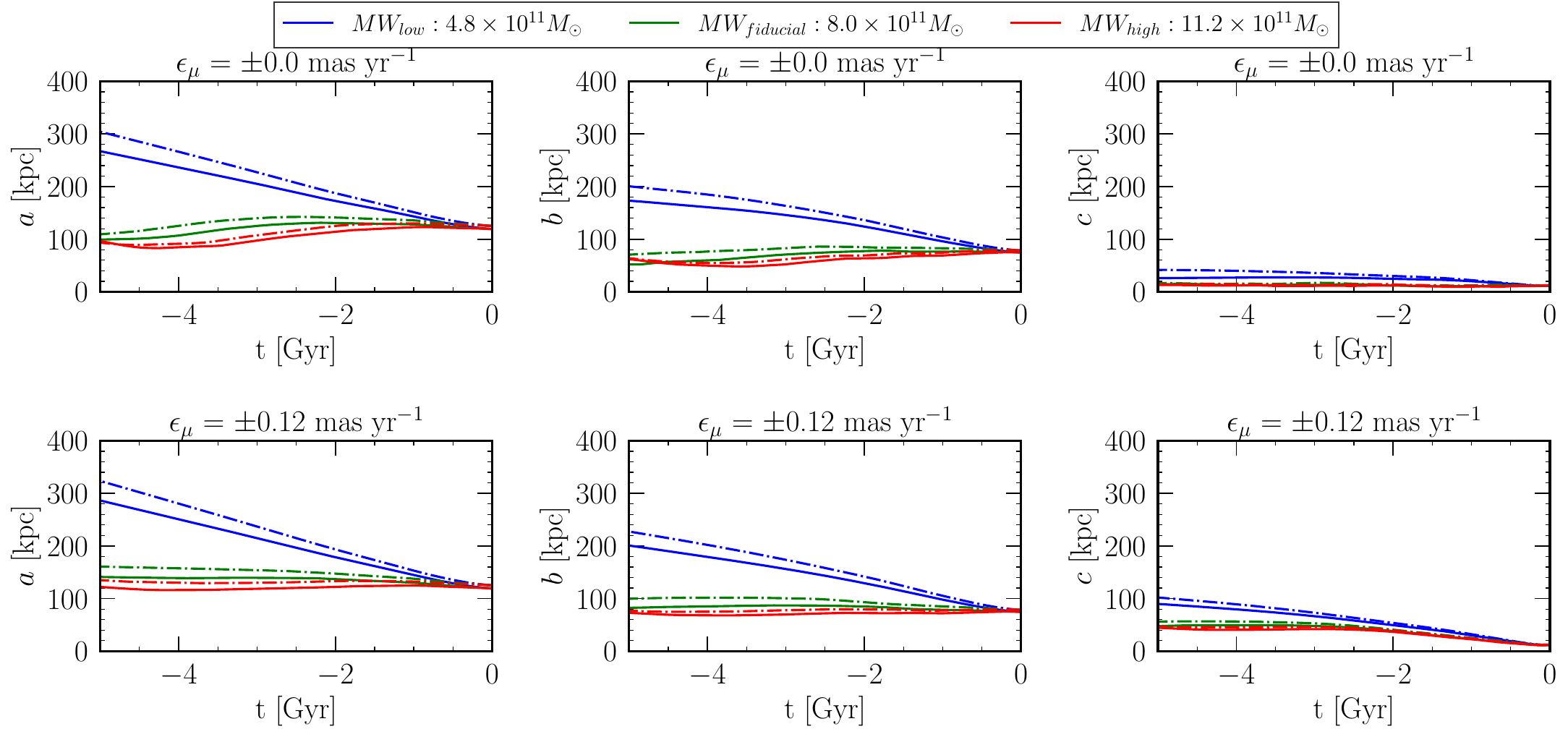}
     \caption{Comparative analysis of galactic orbital axes under proper motion uncertainties and different potential models at $\theta_{tan}$ = 80\textdegree. Panels shows the mean values of 600 major, minor, and intermediate axes calculated through backward integration. Each panel corresponds to a unique combination of proper motion uncertainty and potential models. Dash-dotted curves show the effect of distance uncertainties, while solid curves show results without distance uncertainties. Rows correspond to different levels of proper motion uncertainty, while columns represent the three axes: major (left), intermediate (center), and minor (right).}
     \label{fig:80all_axis_no}
\end{figure*}

The behavior of the minor-to-major axial ratio in the $MW_{\text{low}}$ potential model differs from that in other models, warranting a closer examination of how the extent of the test satellite system changes along its axes, from major to minor. Figure  \ref{fig:80all_axis_no} illustrates the influence of proper motion uncertainties on the mean values of the three axes: $a$ (major), $b$ (intermediate), and $c$ (minor). The influence of proper motion uncertainties is organized by row, while the axes are displayed by column in the figure. Each panel includes curves in three colors, depicting the impact of different potential models. The solid curves represent the results when \(\epsilon_{dist} = 0\%\), while the dashed-dotted curves show the results when \(\epsilon_{dist} = 5\%\). The first row shows $\pm$0.00 mas yr$^{-1}$ proper motion uncertainties and 0\% distance uncertainties. Here, both $MW_{\text{fiducial}}$ and $MW_{\text{high}}$\ show similar behavior across the axes, both with and without distance uncertainties. Over time, the axes tend to decrease, particularly under the $MW_{\text{high}}$ potential model. In contrast, $MW_{\text{low}}$ shows a completely different and opposite behavior. During backward integration, all the axes under $MW_{\text{low}}$ increase over time. 

In these models, the variations in axis lengths are a direct consequence of changes in gravitational pull due to mass differences. In the $MW_{\text{high}}$ model, increased halo mass leads to more compact axes as the structure becomes more gravitationally bound. Conversely, in the $MW_{\text{low}}$ model, a reduction in mass leads to an elongation of the axes, particularly in $a$ and $b$. This suggests that in the $MW_{\text{low}}$ model, more satellites become unbound, causing them to drift away from their host galaxies. We emphasize that the $MW_{\text{fiducial}}$ model serves as a baseline for the $MW_{\text{high}}$ and $MW_{\text{low}}$ models, with the halo mass adjusted by 40\% (increased and decreased, respectively). However, the response of the axis extents is striking, with the Milky Way potential with a 40\% reduction in halo mass having a more pronounced effect on the system than a 40\% increase. Moreover, including distance errors causes each axis in every potential model to extend slightly, indicating a noticeable, though moderate, effect on the orbital axes.

We also tested and recalculated the \( c/a \) ratio by excluding test satellites beyond 300 kpc, considering only those within this threshold, as discussed in Appendix \ref{AppendixPart2_recomputingc_a_300}. We find that this does exclusion not result in a major influence.

\subsection{Inferred escaping satellites}

\begin{figure*}
\centering
   \includegraphics[width=17cm]{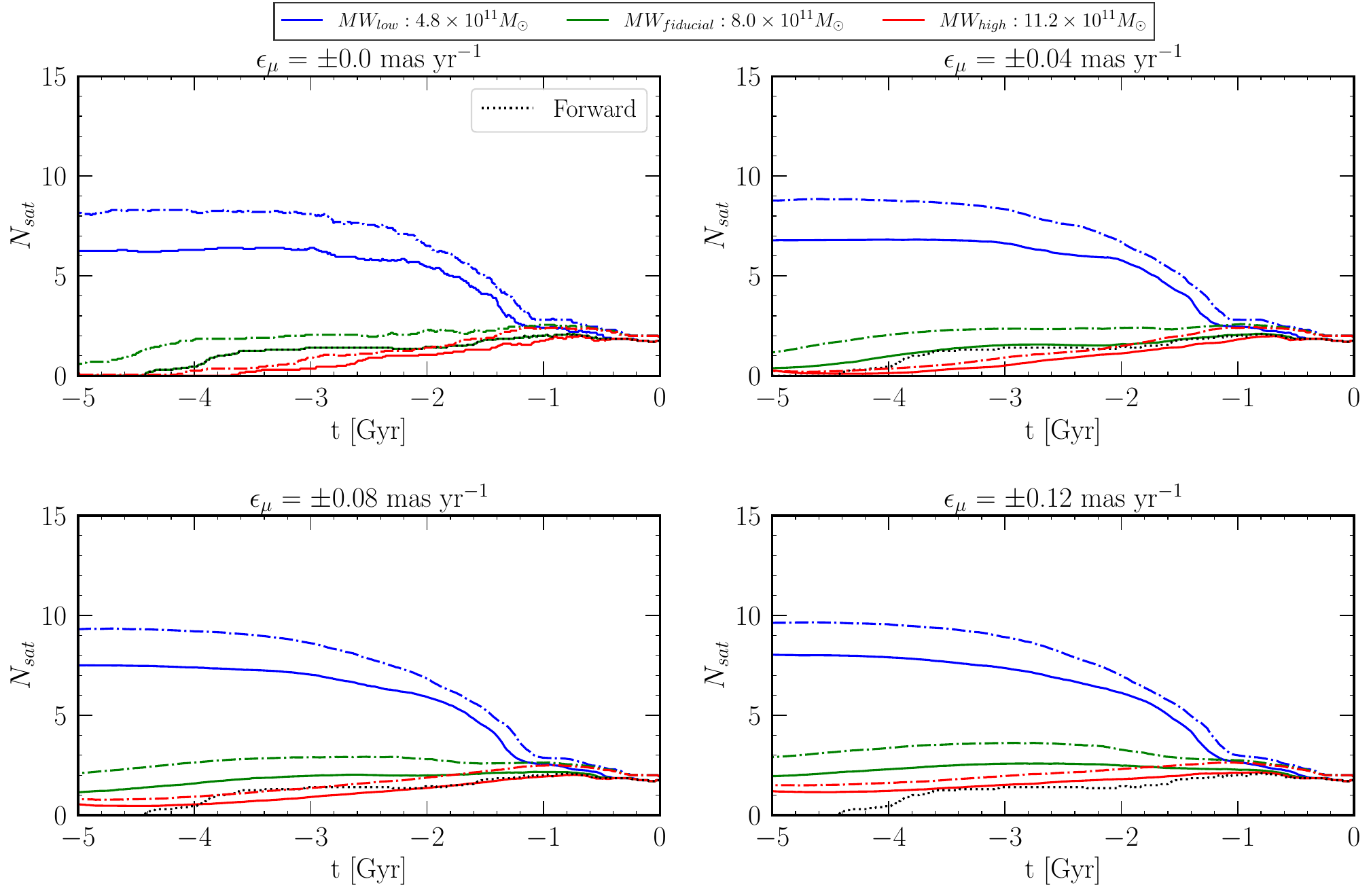}
     \caption{Average number of test satellite exceeding a radial distance of 300 kpc under varying proper motion uncertainties. The four panels correspond to different level of proper motion uncertainty. Each panel  displays six curves, representing the number of test satellites beyond 300 kpc of distance under three different potential models. Dash-dotted curves show results with distance uncertainties included, while solid curves reflect only proper motion errors.}
     \label{fig:unbound_80_no}
\end{figure*}

Figure \ref{fig:unbound_80_no} shows the average number of test satellites that surpass a distance of 300 kpc from the galactic center under proper motion uncertainties and various potential models. The plot is divided into four panels that show three-color curves representing the impact of potential models. The solid curves represent the results for \(\epsilon_{dist} = 0\%\), while the dashed-dotted curves correspond to \(\epsilon_{dist} = 5\%\). In the upper left panel, when there are no proper motion uncertainties, the $MW_{\text{low}}$ curve initially shows a relatively stable number of satellites exceeding 300 kpc in backward integration, both with and without distance uncertainties. Over time, the number gradually increases before reaching a plateau. This outcome is expected, as reducing the halo mass by 40\% lowers the gravitational binding force, allowing more satellites to become unbound and drift beyond 300 kpc.

Unlike the $MW_{\text{low}}$ model, the other two potential models exhibit opposite behavior. In $MW_{\text{fiducial}}$, the number of test satellites that go beyond 300 kpc decreases over time, and at 5 Gyr, all $N_{\text{sat}}$ test satellites remain under the 300 kpc limit. This trend is even more pronounced in the $MW_{\text{high}}$ potential model, where the number of distant test satellites progressively decreases over time due to the higher halo mass. Furthermore, the other three panels demonstrate the effect of proper motion uncertainties on satellite dynamics, showing that as uncertainties increase, more test satellites remain at distances greater than 300 kpc across all models. Proper motion errors preferentially increase the inferred velocities, and thus satellite energies, effectively ejecting more test satellites.

Furthermore, when comparing the curves with and without distance uncertainties, we observe a clear effect: these uncertainties increase the number of test satellites exceeding 300 kpc. In Table \ref{table:4}, the column $N_{\text{sat}}>$ 300 kpc lists models BI-05 to BI-08 without induced distance errors in $MW_{\text{low}}$, and models BI-17 to BI-20, with distance errors. These results indicate that the inclusion of distance errors at 3 Gyr leads to a greater inferred loss of satellites in the system.

\section{Conclusion and outlook}
\label{Sect:conclusion}

We investigated the impact of different types of measurement uncertainties on the inferred stability of artificial, intrinsically stable planes of satellite galaxies. Our key results are as follows.

\begin{itemize}
\item When the test satellite galaxies' proper motions are measured with perfect accuracy (or $\epsilon_{\mu}$ = 0.00 mas yr$^{-1}$) and no other errors are introduced, we infer that the system remains highly stable.

\item At the level of minimum Gaia uncertainties (with $\epsilon_{\mu}$ = 0.04 mas yr$^{-1}$, similar to Gaia systematics), the inferred plane widths begin to diverge from the true value toward larger values. Although the thin initial plane remains relatively well constrained, its inferred width relative to the true plane almost doubles within 3\,Gyr. As we increase the applied proper motion uncertainties ( $\epsilon_{\mu}$ = 0.08 and 0.12 mas yr$^{-1}$), the inferred plane widths increase substantially, and one might infer that the intrinsically stable satellite planes are highly unstable. The $c/a$ shows a linear increase with the addition of more uncertainties.

\item When performing backward integration with three different intrinsic plane heights ($h = 10, 20$, and 30\,kpc), we find that applying realistic uncertainties in the proper motion introduces substantial broadening in the plane of the satellites in all cases. The apparent broadening effect we identified is therefore not sensitive to the exact intrinsic plane height assumed for the plane.

\item When applying 5\% distance uncertainties, we observe a noticeable effect that increases the plane width during backward integration relative to the intrinsic plane width. However, this effect does not dominate the inferred evolution of the satellite plane. Additionally, the impact of distance uncertainties becomes less important with higher levels of proper motion uncertainties.

\item The adoption of incorrect Milky Way potential models leads to an apparent widening of the inferred past satellite planes. This impact is especially significant for $MW_{\text{low}}$, which has a 40\% smaller halo mass compared to the fiducial model considered for our intrinsic satellite planes. Due to its lower mass, $MW_{\text{low}}$ struggles to maintain cohesion among the test satellites, leading to increased radial distances and an extension of its major, minor, and intermediate axes. However, under larger proper motion uncertainties, the influence of adopting incorrect Milky Way potential models diminishes, as the system deviates from its true configuration across all potential models.
\end{itemize}

Thus, this study emphasizes the necessity of precise determination of proper motion, distances of satellite galaxies, and the Milky Way's gravitational potential. Variations in these quantities can significantly shift our inference of the evolution of intrinsically stable satellite planes, especially with respect to proper motion and halo mass of the host galaxy. We argue that underestimating the halo mass also leads to the inference that more satellites escape from the host (and might thus appear unbound) than in the true fiducial model.

Accurate measurement of the Milky Way potential is therefore vital for a comprehensive study of its satellite system and for broader insights into galactic evolution and interactions. We also find that the results are not very sensitive to the number of integrations; in our study, 600 backward integrations per model are sufficient.

However, it is crucial to acknowledge certain limitations in this research. A significant caveat is the omission of the gravitational effects of massive satellite galaxies such as the LMC. These celestial bodies, owing to their substantial mass, can affect the mass estimates of the Milky Way galaxy \citep{erkal2020equilibrium} and exert significant gravitational forces that profoundly influence the Milky Way and its surroundings \citep{garavitocamargo2021,pawlowski2022,vasiliev2023effect}, and thus the stability of the satellite plane. This omission represents a notable limitation in our analysis. Additionally, this study assumes a spherical dark matter halo; however, it is important to acknowledge that the halo could be triaxial \citep{law2009evidence}, which might have implications for our findings.

Future studies could enhance our understanding by incorporating the gravitational influence of massive satellite galaxies, such as the LMC and SMC, as well as by considering different shapes of the dark matter halo. Such an approach would allow for a more comprehensive analysis of the satellite plane dynamics, taking into account the complexities of real-world scenarios. Additionally, in this study, we assume a static dark matter halo over the 5 Gyr integration period. However, this simplification overlooks important evolutionary effects. Over time, the halo mass is expected to grow as it accretes dark matter, resulting in a smaller mass in the past and a larger mass in the future. This affects satellite orbits, particularly at larger radii. \cite{pace2022} focuses on observations and finds that the orbital computations of 46 dwarf spheroidal galaxies (dSphs) using a static Milky Way potential reveal that, while pericenter and apocenter distances can be derived,  the inclusion of the LMC's potential alters these parameters by more than 25\% for 40\% of the sample. Similar studies have also been conducted using cosmological simulations \cite{bell2023uncertainties, santistevan2024modelling}. Additionally, the influence of the baryonic disk leads to adiabatic contraction, increasing the dark matter density at smaller radii, which affects the inner orbits of satellites \citep{cautun2020milky}. However, in general, such additional dynamical effects are expected to increase the uncertainty in past satellite trajectories, as we have only limited constraints on their exact implementation. We therefore expect them to exacerbate the trend of inferring less stable satellite plane dynamics when backward integrating. Moreover, our models does not include an important dynamical process: dynamical friction. This effect, especially for massive satellites orbiting within a host potential, can lead to the gradual decay of orbits as they lose energy and angular momentum to the dark matter background \citep{esquivel2007dynamical, ogiya2016dynamical, taylor2001dynamics}. Over time, this process can significantly alter satellite trajectories and contribute to the disruption of coherent planar structures.

Our method can be extended to examine the observed planes of satellites of M31, Cen A, and others. However, the application to these systems requires careful consideration of observational uncertainties, such as proper motions, distances, and gravitational potentials, which are significantly greater compared to our current analysis (in the case of M31) or thus far inaccessible (for more distant systems). In addition to these observational limitations, the evolution of such planes is also sensitive to the initial distribution of the positions and velocities of the test satellites. Variations in the initial conditions could lead to notable differences in the structure and persistence of the planes over time. By accounting for these factors, our approach can provide valuable insights into the dynamics of satellite galaxies in diverse environments, further broadening the scope of our findings.

Our numerical experiments demonstrate how intrinsically stable, co-orbiting planes of satellite galaxies display a significant increase in their plane height when backward integrated after applying measurement errors. This effect is already evident for proper motion errors comparable to the current Gaia systematics of $\sim 0.033,\mathrm{mas\,yr}^{-1}$ \citep{Lindegren2021,mcconnachie2020updated,li2021gaia}.
\citet{sawala2023milky} argued that such an increase in plane height indicates that the Milky Way's Vast Polar Structure is more likely a transient rather than a rotationally supported feature, and thus consistent with $\Lambda$CDM expectations. Similar arguments were previously put forward by \citet{maji2017there}, based on less accurate proper motion measurements. 
In their studies, the authors integrated the orbits of observed Milky Way satellite galaxies in a static potential, akin to our numerical experiments, and measured the time evolution of the inferred thickness of plane fits. However, our work shows that such an increasing thickness does not necessarily imply an absence of a kinematically coherent, dynamically stable, or long-lived satellite plane. In the case of an underlying, intrinsically stable satellite plane, realistic measurement errors alone, as well as assuming a mismatching Galactic potential, can already cause apparent thickening. Furthermore, care must be taken in how errors are applied. For example, the measured most-likely proper motions of observed satellite galaxies are already subject to measurement error. Sampling the errors in a Monte Carlo fashion (e.g., as demonstrated by \citealt{sawala2023milky}) then effectively applies these errors twice. In our study, this means that our applied proper motion error of $\epsilon_\mu = 0.04\,\mathrm{mas\,yr}^{-1}$\ corresponds to a Monte Carlo sampling of orbits based on observed data with a measurement error of 
$\epsilon_\mu / \sqrt{2} = 0.028\,\mathrm{mas\,yr}^{-1}$. Monte Carlo sampling therefore enhances the effect we identified in our study of artificially increasing the inferred satellite plane height when backward integrating, because it typically adds more dispersion to an ensemble of satellites.
Merely finding an increase in the inferred plane height of an observed satellite structure under orbit integration--without accounting for the impact of measurement errors (especially in proper motions) or mismatches in the assumed Galactic gravitational potential--is therefore insufficient to claim that the structure is not dynamically stable or transient, or that it is therefore consistent with $\Lambda$CDM.

\begin{acknowledgements}
MSP acknowledges funding via a Leibniz-Junior Research Group (project number J94/2020). Moreover, this work made use of the following software packages: \texttt{astropy} \citep{astropy:2013, astropy:2018, astropy:2022}, \texttt{matplotlib} \citep{Hunter:2007}, \texttt{numpy} \citep{numpy}, \texttt{python} \citep{python}, \texttt{scipy} \citep{2020SciPy-NMeth, scipy_14880408}, and \texttt{galpy} \citep{bovy2015galpy}.  Software citation information aggregated using \texttt{\href{https://www.tomwagg.com/software-citation-station/}{The Software Citation Station}} \citep{software-citation-station-paper, software-citation-station-zenodo}.
\end{acknowledgements}

\bibliographystyle{aa}
\bibliography{ref}

\begin{appendix}

\section{Analysis at $\theta_{\max}=40^\circ$ and $60^\circ$}

\label{Appendix}

\begin{table*}[b]

\caption{Parameters and results for backward integration models at 3 Gyr in backward integration with \(\theta_{max} = 80^\circ\).}
\label{table:4} 
\centering
\begin{tabular}{ccccccccc}
\hline\hline

Model & $\epsilon_{\mu}$ & $\epsilon_{dist}$ & \multicolumn{2}{c}{Potential} & M$_{\text{realization}}$ & $\Delta_{c/a}$ & $f_{c/a}$ & $N_{\text{sat}}(> 300 \text{ kpc})$ \\
 & [mas yr$^{-1}$] & [\%] & Forward & Backward & & & & \\
\midrule

BI-01 &                      $\pm$ 0.00  &                       0  & $MW_{\text{fiducial}}$ &       $MW_{\text{fiducial}}$  &    600 &         0.00 $\pm$ 0.01  & 1.00 $\pm$ 0.12  &     1.40 $\pm$ 0.92 \\
BI-02 &                      $\pm$ 0.04  &                       0  &  $MW_{\text{fiducial}}$ &      $MW_{\text{fiducial}}$  &         600 &    0.07 $\pm$ 0.05  & 1.69 $\pm$ 0.51  &     1.53 $\pm$ 1.01 \\
BI-03 &                      $\pm$ 0.08  &                       0  &   $MW_{\text{fiducial}}$ &     $MW_{\text{fiducial}}$  &       600 &      0.17 $\pm$ 0.09  & 2.64 $\pm$ 0.87  &     1.97 $\pm$ 1.10 \\
BI-04 &                      $\pm$ 0.12  &                       0  &   $MW_{\text{fiducial}}$ &     $MW_{\text{fiducial}}$  &      600 &       0.24 $\pm$ 0.11  & 3.32 $\pm$ 1.07  &     2.57 $\pm$ 1.14 \\
\hline
BI-05 &                      $\pm$ 0.00  &                       0  &    $MW_{\text{fiducial}}$ &        $MW_{\text{low}}$  &       600 &      0.03 $\pm$ 0.05  & 1.27 $\pm$ 0.44  &     6.40 $\pm$ 1.50 \\
BI-06 &                      $\pm$ 0.04  &                       0  &   $MW_{\text{fiducial}}$ &         $MW_{\text{low}}$  &      600 &       0.06 $\pm$ 0.05  & 1.61 $\pm$ 0.47  &     6.63 $\pm$ 1.46 \\
BI-07 &                      $\pm$ 0.08  &                       0  &   $MW_{\text{fiducial}}$ &         $MW_{\text{low}}$  &      600 &       0.13 $\pm$ 0.07  & 2.26 $\pm$ 0.68  &     7.03 $\pm$ 1.38 \\
BI-08 &                      $\pm$ 0.12  &                       0  &  $MW_{\text{fiducial}}$ &          $MW_{\text{low}}$  &      600 &       0.21 $\pm$ 0.10  & 2.97 $\pm$ 0.93  &     7.36 $\pm$ 1.43 \\
\hline
BI-09 &                      $\pm$ 0.00  &                       0  & $MW_{\text{fiducial}}$ &           $MW_{\text{high}}$  &       600 &      0.01 $\pm$ 0.02  & 1.13 $\pm$ 0.16  &     0.35 $\pm$ 0.48 \\
BI-10 &                      $\pm$ 0.04  &                       0  &  $MW_{\text{fiducial}}$ &          $MW_{\text{high}}$  &      600 &       0.10 $\pm$ 0.06  & 1.93 $\pm$ 0.58  &     0.51 $\pm$ 0.59 \\
BI-11 &                      $\pm$ 0.08  &                       0  &  $MW_{\text{fiducial}}$ &          $MW_{\text{high}}$  &      600 &       0.21 $\pm$ 0.10  & 2.98 $\pm$ 0.97  &     0.91 $\pm$ 0.79 \\
BI-12 &                      $\pm$ 0.12  &                       0  &  $MW_{\text{fiducial}}$ &          $MW_{\text{high}}$  &      600 &       0.26 $\pm$ 0.11  & 3.45 $\pm$ 1.02  &     1.51 $\pm$ 0.93 \\
\hline
BI-13 &                      $\pm$ 0.00  &                       5  &   $MW_{\text{fiducial}}$ &     $MW_{\text{fiducial}}$  &      600 &       0.02 $\pm$ 0.02  & 1.17 $\pm$ 0.18  &     2.05 $\pm$ 1.07 \\
BI-14 &                      $\pm$ 0.04  &                       5  &   $MW_{\text{fiducial}}$ &     $MW_{\text{fiducial}}$  &      600 &       0.08 $\pm$ 0.05  & 1.78 $\pm$ 0.51  &     2.35 $\pm$ 1.12 \\
BI-15 &                      $\pm$ 0.08  &                       5  &   $MW_{\text{fiducial}}$ &     $MW_{\text{fiducial}}$  &     600 &        0.18 $\pm$ 0.09   & 2.68 $\pm$ 0.86  &     2.90 $\pm$ 1.14 \\
BI-16 &                      $\pm$ 0.12  &                       5  &   $MW_{\text{fiducial}}$ &     $MW_{\text{fiducial}}$  &     600 &        0.24 $\pm$ 0.11  & 3.32 $\pm$ 1.06  &     3.61 $\pm$ 1.17 \\
\hline
BI-17 &                      $\pm$ 0.00  &                       5  &    $MW_{\text{fiducial}}$ &        $MW_{\text{low}}$  &      600 &       0.05 $\pm$ 0.06   & 1.51 $\pm$ 0.57  &     8.10 $\pm$ 1.58 \\
BI-18 &                      $\pm$ 0.04  &                       5  &    $MW_{\text{fiducial}}$ &        $MW_{\text{low}}$  &      600 &       0.08 $\pm$ 0.06   & 1.80 $\pm$ 0.57   &     8.34 $\pm$ 1.52 \\
BI-19 &                      $\pm$ 0.08  &                       5  &    $MW_{\text{fiducial}}$ &        $MW_{\text{low}}$  &      600 &       0.15 $\pm$ 0.08  & 2.41 $\pm$ 0.71  &     8.61 $\pm$ 1.53 \\
BI-20 &                      $\pm$ 0.12  &                       5  &   $MW_{\text{fiducial}}$ &         $MW_{\text{low}}$  &      600 &       0.21 $\pm$ 0.08  & 2.95 $\pm$ 0.79   &     8.92 $\pm$ 1.54 \\
\hline
BI-21 &                      $\pm$ 0.00  &                       5  &    $MW_{\text{fiducial}}$ &        $MW_{\text{high}}$  &    600 &         0.02 $\pm$ 0.02  & 1.16 $\pm$ 0.21  &     0.90 $\pm$ 0.77 \\
BI-22 &                      $\pm$ 0.04  &                       5  &   $MW_{\text{fiducial}}$ &         $MW_{\text{high}}$  &      600 &       0.10 $\pm$ 0.06  & 1.94 $\pm$ 0.61  &     0.92 $\pm$ 0.74 \\
BI-23 &                      $\pm$ 0.08  &                       5  &    $MW_{\text{fiducial}}$ &        $MW_{\text{high}}$  &     600 &        0.21 $\pm$ 0.10  & 2.97 $\pm$ 0.97  &     1.36 $\pm$ 0.91 \\
BI-24 &                      $\pm$ 0.12  &                       5  &   $MW_{\text{fiducial}}$ &         $MW_{\text{high}}$  &     600 &        0.26 $\pm$ 0.11  & 3.46 $\pm$ 1.05  &     1.97 $\pm$ 1.02 \\

\bottomrule

\end{tabular}
\tablefoot{The table lists the model numbers (BI), proper motion errors (\(\epsilon_\mu\)) in milliarcseconds per year, distance errors (\(\epsilon_{dist}\)) in percentage, the gravitational potential used for forward and backward integration (\(\textit{MW}_{\text{fiducial}}\), \(\textit{MW}_{\text{low}}\), \(\textit{MW}_{\text{high}}\)), the number of realizations M$_{\text{realization}}$ for each backward integration, the average absolute difference in average minor-to-major axis ratio flattening between forward and backward integration (\(\Delta_{c/a}\)), and the average relative change in this quantity between the forward and backward integration (\(f_{c/a}\)) along with their standard deviations (std) as a measure of their spread between different realizations at 3 Gyr. Additionally, \(N_{sat} > 300 \, \text{kpc}\) denotes the absolute number of satellite realizations beyond 300 kpc with their standard deviations.}
\end{table*}

This appendix presents the simulations involving less eccentric orbits. Results and analysis for  $\theta_{tan} = 80\%$ have been discussed in detail in 
Section \ref{Sect:Results}. Following a similar pattern, we here present the results which are obtained for  $\theta_{tan}$ = 40\textdegree and 60\textdegree.

\begin{figure*}[t]
\centering
   \includegraphics[width=17cm]{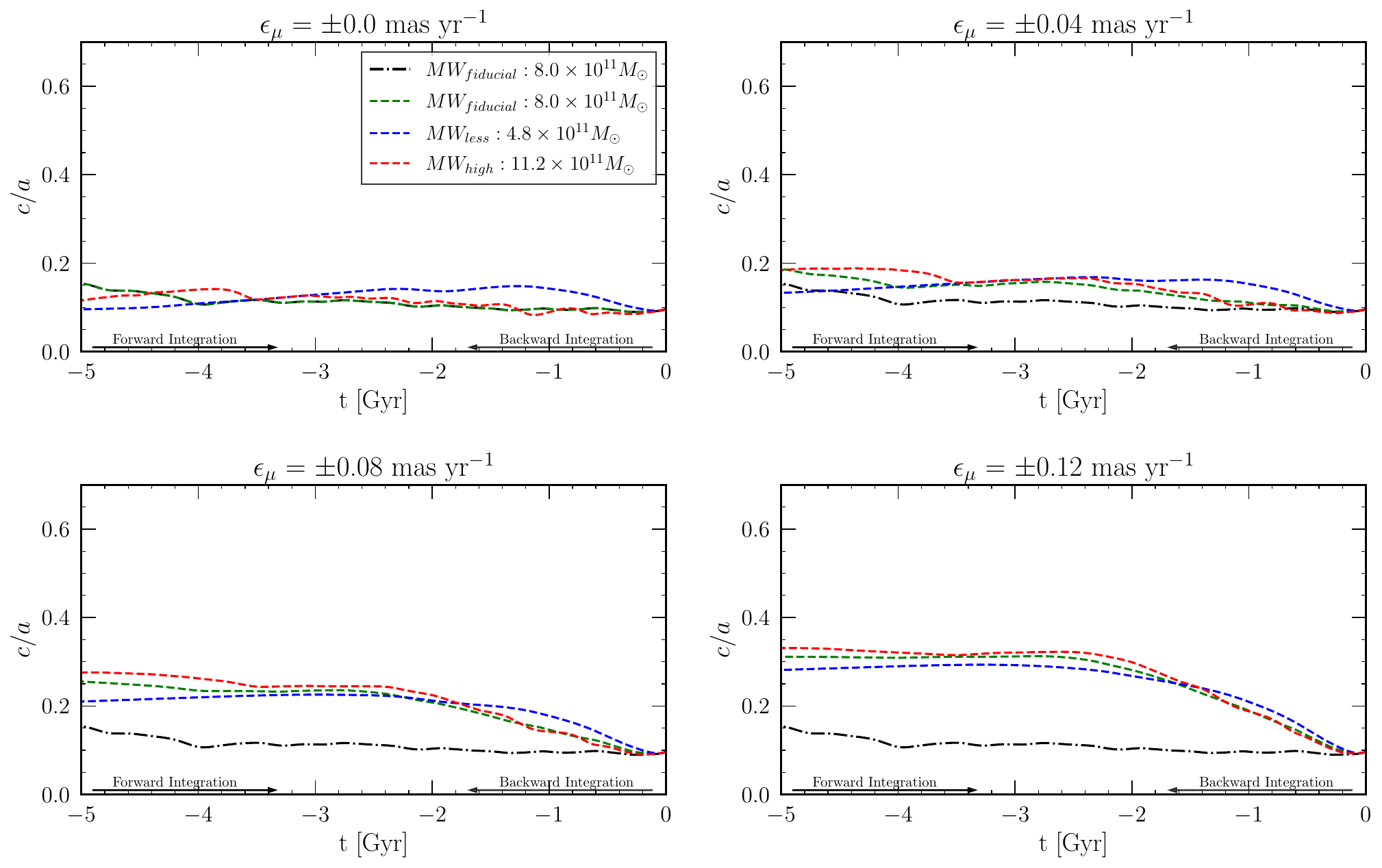}
     \caption{Analogous to Fig \ref{fig:80_all_mass_all_pmra_no}, the results for $\theta_{tan}$= 40\textdegree: Evolution of the plane of satellite galaxies with proper motion uncertainties included in each panel. The black dashed-dotted curves represent the mean axis ratio for forward integration, while the colored dashed curves indicate the mean axis ratio for backward integration under different potential models characterized by halo mass. Specifically, the blue, green, and red dashed curves correspond to the ${c/a}$ values under the $MW_{\text{low}}$, $MW_{\text{fiducial}}$, and $MW_{\text{high}}$ potential models, respectively. Note that these results do not account for distance uncertainties.}
     \label{fig:40_all_mass_all_pmra_no}
\end{figure*}

\begin{figure*}
\centering
   \includegraphics[width=15cm]{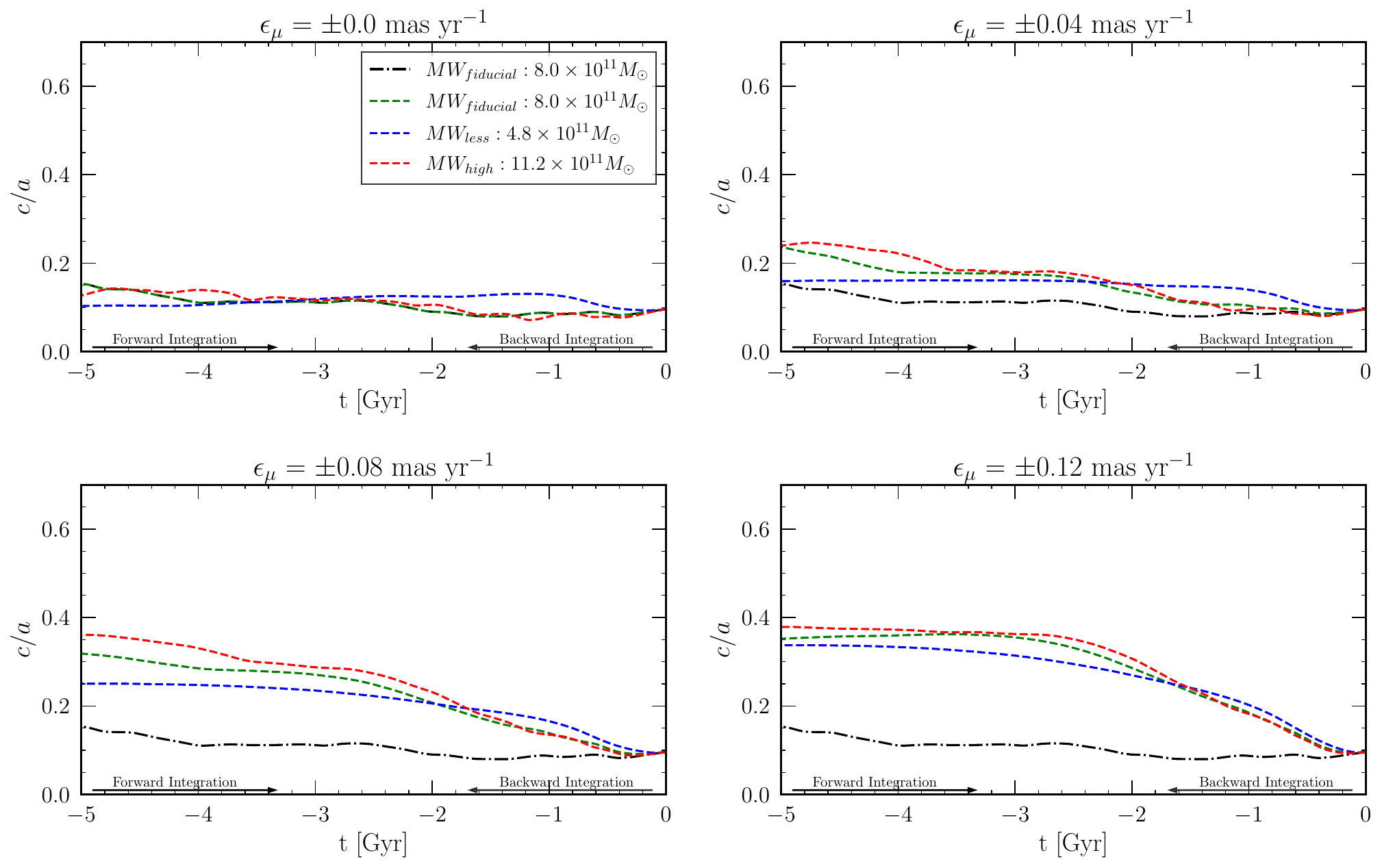}
     \caption{Analogous to Fig \ref{fig:80_all_mass_all_pmra_no}, results for $\theta_{tan}$= 60\textdegree: Evolution of the plane of satellite galaxies with proper motion uncertainties included in each panel. The black dashed-dotted curves represent the mean axis ratio for forward integration, while the colored dashed curves indicate the mean axis ratio for backward integration under different potential models characterized by halo mass. Specifically, the blue, green, and red dashed curves correspond to the ${c/a}$ values under the $MW_{\text{low}}$, $MW_{\text{fiducial}}$, and $MW_{\text{high}}$ potential models, respectively. Note that these results do not account for distance uncertainties.}
     \label{fig:60_all_mass_all_pmra_no}
\end{figure*}

Figure \ref{fig:40_all_mass_all_pmra_no} and  \ref{fig:60_all_mass_all_pmra_no} show results analogous to those in Section \ref{Sect:Results}. These scenarios exhibit qualitatively comparable patterns. The figures are segmented into four panels, each representing different levels of proper motion uncertainties. The mean $c/a$ ratios for forward integration are depicted with black curves, whereas those for backward integration are shown in colored curves. Similar to the previously discussed case, it becomes increasingly difficult to accurately determine the stability of the satellite plane as the level of proper motion errors escalates. 
Furthermore, in these results too, the behavior of the system under $MW_{\text{low}}$ is different from the other two potential models. 

Figure \ref{fig:e} shows that as \(\theta_{\text{max}}\) increases, the orbits become more eccentric, and this effect is clearly observed when we consider the results under \(\theta_{\text{max}}\). By comparing the results at \(\theta_{\text{max}} = 40\%, 60\%, 80\%\), as shown in Tables \ref{table:deg_40}, \ref{table:deg_60}, and \ref{table:4}, we see that at a given level of proper motion uncertainty, the satellite plane widens more for larger \(\theta_{\text{max}}\). For instance, in Model BI-02, when \(\epsilon_{\mu} = 0.04\) mas yr\(^{-1}\), \(\Delta_{c/a}\) is 0.04, 0.06, and 0.07 for \(\theta_{\text{tan}} = 40\%, 60\%, 80\%\), respectively, and increases gradually. Furthermore, the influence of eccentric orbits is also apparent when comparing the curves in Fig. \ref{fig:40_all_mass_all_pmra_no}, \ref{fig:60_all_mass_all_pmra_no}, and \ref{fig:80_all_mass_all_pmra_no}. In cases with less eccentric orbits (\(\theta_{\text{tan}} = 40\%\)), the plane appears relatively smoother and flatter. In contrast, as the orbits become more eccentric, the evolution of the plane flattening becomes increasingly irregular and pronounced, indicating a significant impact of orbital eccentricity on the satellite's dynamics. This suggests that as \(\theta_{\text{max}}\) increases, not only does the eccentricity of the orbits intensify, but the overall structure and stability of the orbital plane are also affected, leading to a more complex and varied orbital configuration.

\begin{figure*}
\centering
   \includegraphics[width=15cm]{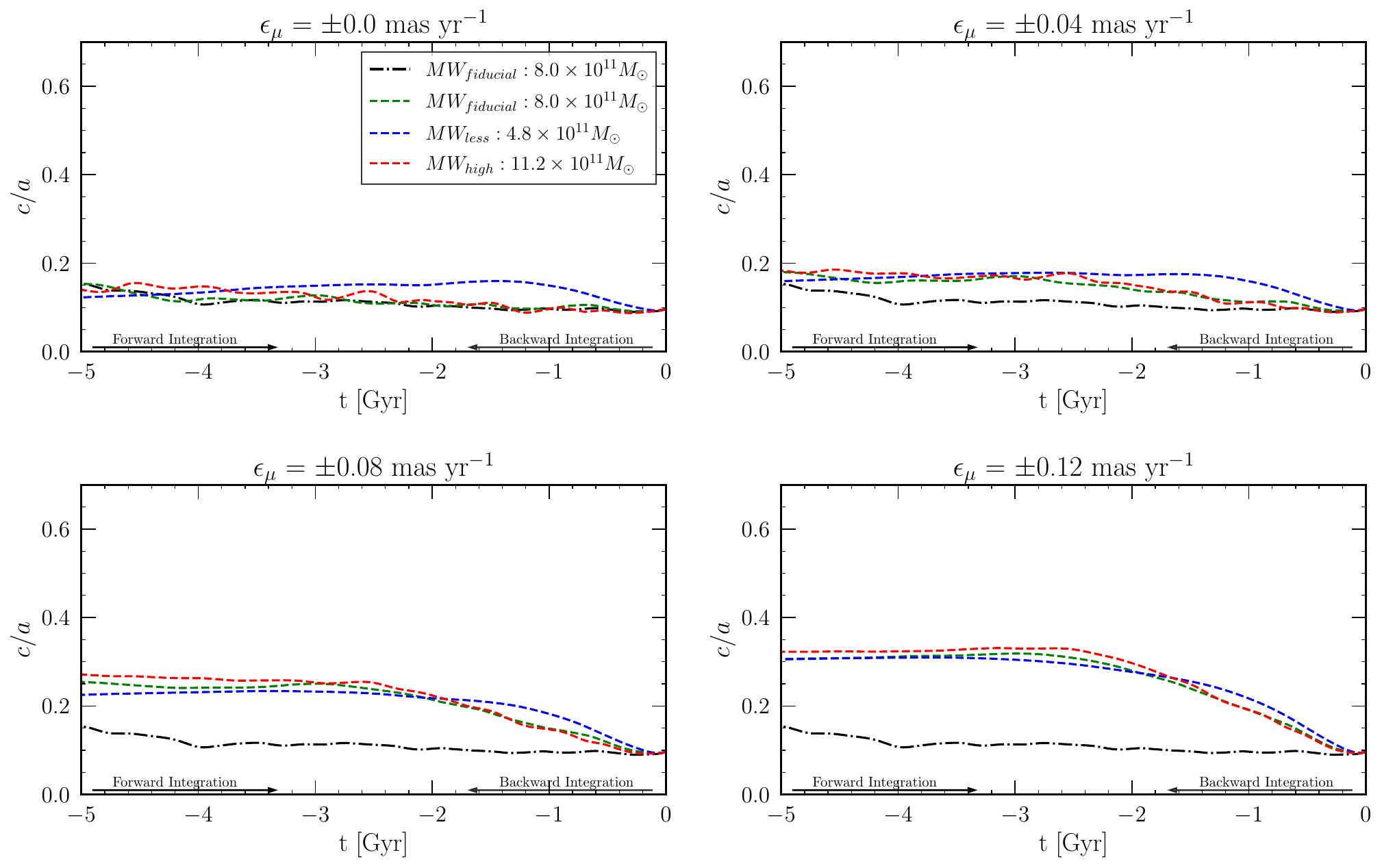}
    \caption{Analogous to Fig \ref{fig:80_all_mass_all_pmra_yes}, results for $\theta_{tan}$= 40\textdegree: Evolution of the plane of satellite galaxies with proper motion uncertainties included in each panel. The black dashed-dotted curves represent the mean axis ratio for forward integration, while the colored dashed curves show the mean axis ratio for backward integration under different potential models characterized by halo mass. Specifically, the blue, green, and red dashed curves correspond to the ${c/a}$ values under the $MW_{\text{low}}$, $MW_{\text{fiducial}}$, and $MW_{\text{high}}$ potential models, respectively. Note that in these results we account  $5\%$ of distance uncertainties.}

     \label{fig:40_all_mass_all_pmra_yes}
\end{figure*}

\begin{figure*}
\centering
   \includegraphics[width=15cm]{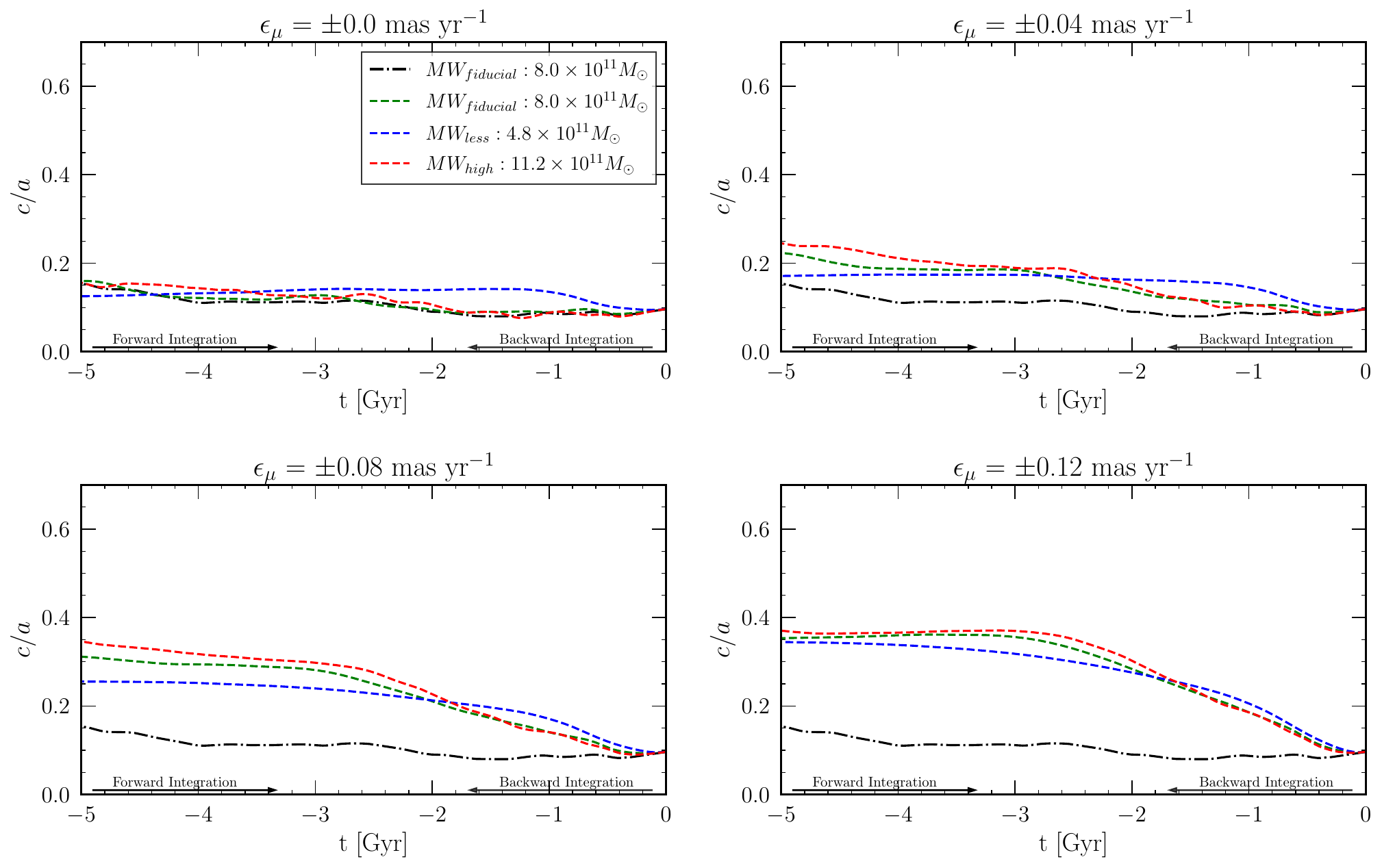}
    \caption{Analogous to Fig \ref{fig:80_all_mass_all_pmra_yes}, results for $\theta_{tan}$= 60\textdegree: Evolution of the plane of satellite galaxies with proper motion uncertainties included in each panel. The black dashed-dotted curves represent the mean axis ratio for forward integration, while the colored dashed curves indicate the mean axis ratio for backward integration under different potential models characterized by halo mass. Specifically, the blue, green, and red dashed curves correspond to the ${c/a}$ values under the $MW_{\text{low}}$, $MW_{\text{fiducial}}$, and $MW_{\text{high}}$ potential models, respectively. Note that in these results we account  $5\%$ of distance uncertainties.}

     \label{fig:60_all_mass_all_pmra_yes}
\end{figure*}

Similarly, Fig. \ref{fig:40_all_mass_all_pmra_yes} and \ref{fig:60_all_mass_all_pmra_yes} display the effects of distance uncertainties on the test satellite system for $\theta_{tan}$= 40\textdegree and 60\textdegree, respectively. The analysis, as discussed in Section \ref{Sect:Results}, indicates that the inclusion of distance errors has a noticeable effect on the $c/a$. However, this does not significantly impact the overall evolution of  $c/a$.

\begin{figure*}
\centering
   \includegraphics[width=15cm]{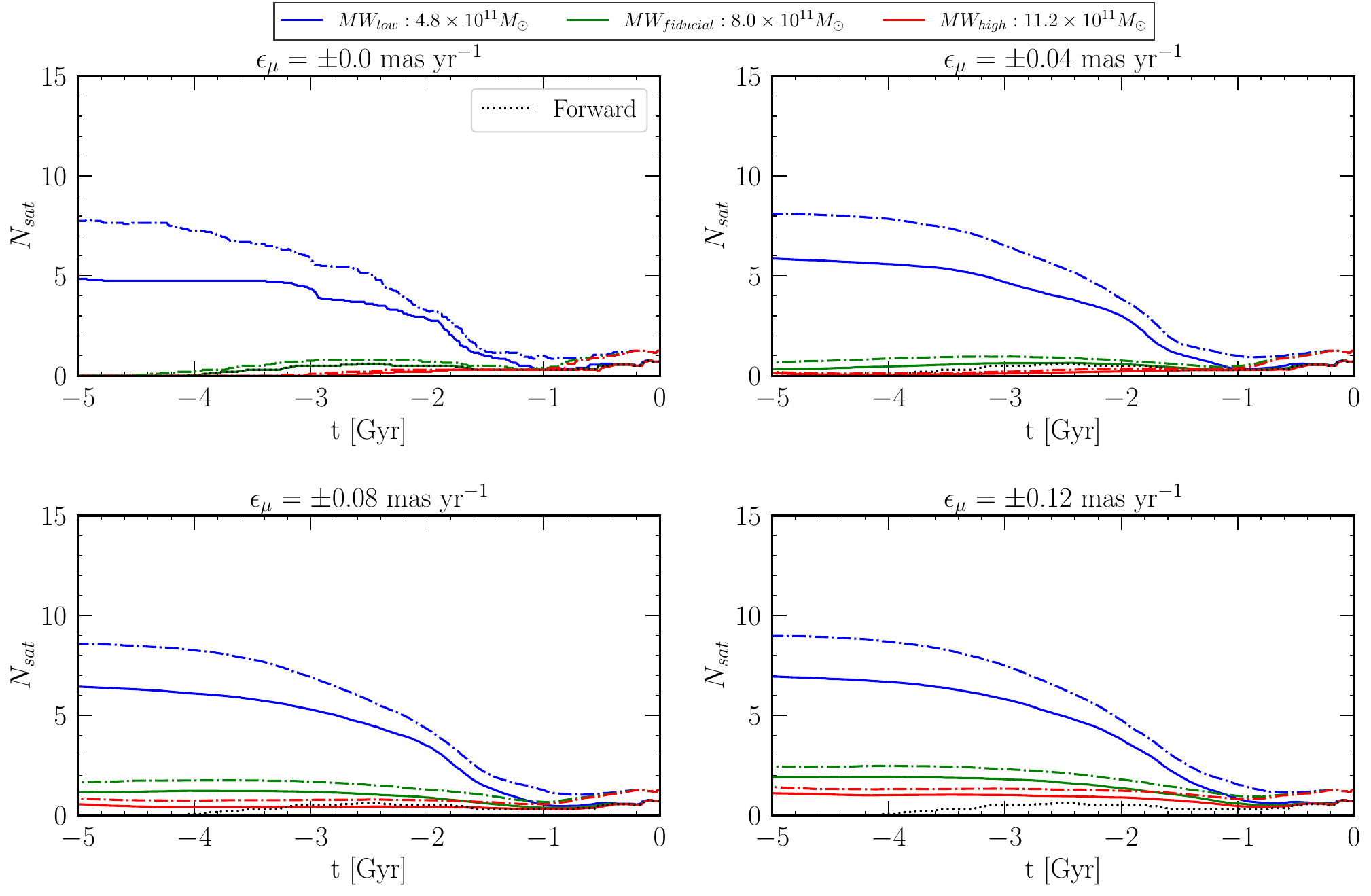}
     \caption{Average number of test satellite exceeding 300 kpc of radial distance under proper motion uncertainties. Similar to Fig. \ref{fig:unbound_80_no}, but under $\theta_{tan}$= 40\textdegree.}
     \label{fig:unbound_40_both}
\end{figure*}

\begin{figure*}
\centering
   \includegraphics[width=15cm]{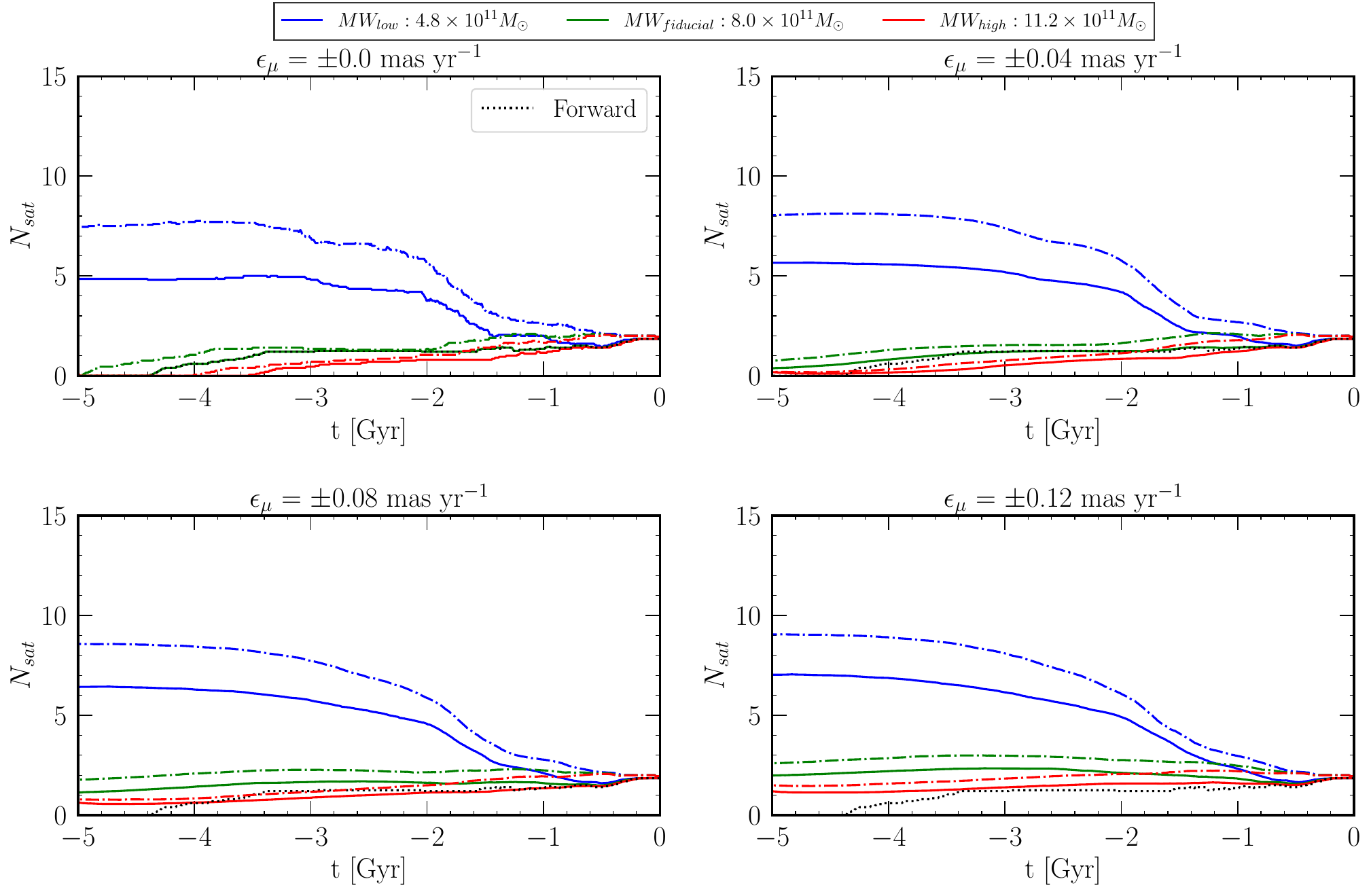}
     \caption{Average number of test satellite exceeding 300 kpc of radial distance under proper motion uncertainties. Similar to Fig. \ref{fig:unbound_80_no}, but under $\theta_{tan}$= 60\textdegree.}
     \label{fig:unbound_60_both}
\end{figure*}

Finally, similar to Fig. \ref{fig:unbound_80_no}, here Fig. \ref{fig:unbound_40_both} and \ref{fig:unbound_60_both} display the number of test satellites that exceeded the 300 kpc radius threshold under $\theta_{max}$ = 40\textdegree and 60\textdegree, respectively. The effect of proper motion errors is depicted in each of the four panels, with each line within a panel representing the influence of different Milky Way potential models. Dash-dotted curves show the effect of 5\% distance uncertainties, while solid curves show 0\% distance uncertainties. The influence of eccentricities is clearly apparent in these figures. For a given model, less eccentric orbits result in fewer test satellites escaping the system compared to those with higher eccentricities. This trend is further corroborated by the data presented in Tables \ref{table:4}, \ref{table:deg_40}, and \ref{table:deg_60}. In these tables, under all models, the column representing the number of satellites with distances greater than 300 kpc ($N_{sat}$ $>$ 300 kpc) consistently increases as the eccentricity of the orbits increases.

\begin{table*}[b]

\caption{Parameters and results for backward integration models at 3 Gyr in backward integration, similar to Table \ref{table:4} but at \(\theta_{max} = 40^\circ\).}
\label{table:deg_40} 
\centering

\begin{tabular}{ccccccccc}
\hline\hline

 Model & $\epsilon_{\mu}$ & $\epsilon_{dist}$ & \multicolumn{2}{c}{Potential} & M$_{\text{realization}}$ & $\Delta_{c/a}$ & $f_{c/a}$ & $N_{\text{sat}}(> 300 \text{ kpc})$ \\
 & [mas yr$^{-1}$] & [\%] & Forward & Backward & & & & \\
\midrule

BI-01 &                      $\pm$ 0.00  &                       0  &   $MW_{\text{fiducial}}$ &    $MW_{\text{fiducial}}$  &     600 &        0.00 $\pm$ 0.02  & 1.00 $\pm$ 0.18  &     0.50 $\pm$ 0.59 \\
BI-02 &                      $\pm$ 0.04  &                       0  &   $MW_{\text{fiducial}}$ &       $MW_{\text{fiducial}}$  &      600 &       0.04 $\pm$ 0.04  & 1.37 $\pm$ 0.34  &     0.63 $\pm$ 0.72 \\
BI-03 &                      $\pm$ 0.08  &                       0  &   $MW_{\text{fiducial}}$ &       $MW_{\text{fiducial}}$  &     600 &        0.12 $\pm$ 0.08  & 2.07 $\pm$ 0.69  &     1.17 $\pm$ 0.94 \\
BI-04 &                      $\pm$ 0.12  &                       0  &   $MW_{\text{fiducial}}$ &       $MW_{\text{fiducial}}$  &    600 &         0.20 $\pm$ 0.12  & 2.75 $\pm$ 1.04  &     1.80 $\pm$ 1.15 \\
\hline
BI-05 &                      $\pm$ 0.00  &                       0  &   $MW_{\text{fiducial}}$ &           $MW_{\text{low}}$  &     600 &        0.02 $\pm$ 0.03  & 1.13 $\pm$ 0.28  &     4.35 $\pm$ 1.82 \\
BI-06 &                      $\pm$ 0.04  &                       0  &   $MW_{\text{fiducial}}$ &           $MW_{\text{low}}$  &    600 &         0.05 $\pm$ 0.04  & 1.43 $\pm$ 0.35  &     4.69 $\pm$ 1.65 \\
BI-07 &                      $\pm$ 0.08  &                       0  &    $MW_{\text{fiducial}}$ &          $MW_{\text{low}}$  &    600 &         0.11 $\pm$ 0.07  & 1.99 $\pm$ 0.59  &     5.29 $\pm$ 1.61 \\
BI-08 &                      $\pm$ 0.12  &                       0  &    $MW_{\text{fiducial}}$ &          $MW_{\text{low}}$  &    600 &         0.18 $\pm$ 0.10  & 2.58 $\pm$ 0.84  &     5.80 $\pm$ 1.67 \\
\hline
BI-09 &                      $\pm$ 0.00  &                       0  &   $MW_{\text{fiducial}}$ &           $MW_{\text{high}}$  &   600 &          0.01 $\pm$ 0.02  & 1.11 $\pm$ 0.16  &     0.00 $\pm$ 0.00 \\
BI-10 &                      $\pm$ 0.04  &                       0  &   $MW_{\text{fiducial}}$ &           $MW_{\text{high}}$  &   600 &          0.05 $\pm$ 0.04  & 1.43 $\pm$ 0.31  &     0.12 $\pm$ 0.33 \\
BI-11 &                      $\pm$ 0.08  &                       0  &   $MW_{\text{fiducial}}$ &           $MW_{\text{high}}$  &   600 &          0.13 $\pm$ 0.08  & 2.16 $\pm$ 0.72  &     0.43 $\pm$ 0.61 \\
BI-12 &                      $\pm$ 0.12  &                       0  &   $MW_{\text{fiducial}}$ &           $MW_{\text{high}}$  &    600 &         0.21 $\pm$ 0.11  & 2.83 $\pm$ 0.10  &     1.01 $\pm$ 0.88 \\
\hline
BI-13 &                      $\pm$ 0.00  &                       5  &   $MW_{\text{fiducial}}$ &       $MW_{\text{fiducial}}$  &     600 &        0.01 $\pm$ 0.02  & 1.12 $\pm$ 0.145  &    0.75 $\pm$ 0.83 \\
BI-14 &                      $\pm$ 0.04  &                       5  &    $MW_{\text{fiducial}}$ &      $MW_{\text{fiducial}}$  &    600 &         0.06 $\pm$ 0.04  & 1.51 $\pm$ 0.37  &     0.97 $\pm$ 0.85 \\
BI-15 &                      $\pm$ 0.08  &                       5  &   $MW_{\text{fiducial}}$ &       $MW_{\text{fiducial}}$  &    600 &         0.14 $\pm$ 0.08  & 2.21 $\pm$ 0.73  &     1.67 $\pm$ 1.05 \\
BI-16 &                      $\pm$ 0.12  &                       5  &    $MW_{\text{fiducial}}$ &      $MW_{\text{fiducial}}$  &   600 &          0.21 $\pm$ 0.11  & 2.81 $\pm$ 0.93  &     2.31 $\pm$ 1.24 \\
\hline
BI-17 &                      $\pm$ 0.00  &                       5  &   $MW_{\text{fiducial}}$ &           $MW_{\text{low}}$  &     600 &        0.04 $\pm$ 0.04  & 1.31 $\pm$ 0.36  &     5.80 $\pm$ 2.18 \\
BI-18 &                      $\pm$ 0.04  &                       5  &    $MW_{\text{fiducial}}$ &          $MW_{\text{low}}$  &    600 &         0.06 $\pm$ 0.04  & 1.57 $\pm$ 0.39  &     6.51 $\pm$ 1.90 \\
BI-19 &                      $\pm$ 0.08  &                       5  &    $MW_{\text{fiducial}}$ &          $MW_{\text{low}}$  &    600 &         0.12 $\pm$ 0.06  & 2.05 $\pm$ 0.54  &     6.92 $\pm$ 1.87 \\
BI-20 &                      $\pm$ 0.12  &                       5  &    $MW_{\text{fiducial}}$ &          $MW_{\text{low}}$  &     600 &        0.19 $\pm$ 0.09  & 2.69 $\pm$ 0.77  &     7.48 $\pm$ 1.91 \\
\hline
BI-21 &                      $\pm$ 0.00  &                       5  &    $MW_{\text{fiducial}}$ &          $MW_{\text{high}}$  &    600 &         0.01 $\pm$ 0.02  & 1.09 $\pm$ 0.14  &     0.10 $\pm$ 0.30 \\
BI-22 &                      $\pm$ 0.04  &                       5  &   $MW_{\text{fiducial}}$ &           $MW_{\text{high}}$  &    600 &         0.05 $\pm$ 0.04  & 1.46 $\pm$ 0.36  &     0.21 $\pm$ 0.44 \\
BI-23 &                      $\pm$ 0.08  &                       5  &    $MW_{\text{fiducial}}$ &          $MW_{\text{high}}$  &    600 &         0.14 $\pm$ 0.09  & 2.23 $\pm$ 0.80  &     0.77 $\pm$ 0.78 \\
BI-24 &                      $\pm$ 0.12  &                       5  &   $MW_{\text{fiducial}}$ &           $MW_{\text{high}}$  &      600 &       0.22 $\pm$ 0.12  & 2.91 $\pm$ 1.07  &     1.32 $\pm$ 0.94 \\

\hline
\end{tabular}
\end{table*}

\begin{table*}[b]
\caption{Parameters and results for backward integration models at 3 Gyr in backward integration, similar to Table \ref{table:4} but at \(\theta_{max} = 60^\circ\)}.
\label{table:deg_60} 
\centering
\begin{tabular}{ccccccccc}
\hline\hline
 Model & $\epsilon_{\mu}$ & $\epsilon_{dist}$ & \multicolumn{2}{c}{Potential} & M$_{\text{realization}}$ & $\Delta_{c/a}$ & $f_{c/a}$ & $N_{\text{sat}}(> 300 \text{ kpc})$ \\
 & [mas yr$^{-1}$] & [\%] & Forward & Backward & & & & \\
\midrule

BI-01 &                      $\pm$ 0.00  &                       0  &    $MW_{\text{fiducial}}$  &     $MW_{\text{fiducial}}$  &   600 &          0.00 $\pm$ 0.01  & 1.00 $\pm$ 0.10  &     1.20 $\pm$ 0.81 \\
BI-02 &                      $\pm$ 0.04  &                       0  &    $MW_{\text{fiducial}}$  &     $MW_{\text{fiducial}}$  &    600 &         0.06 $\pm$ 0.04  & 1.57 $\pm$ 0.39  &     1.19 $\pm$ 0.83 \\
BI-03 &                      $\pm$ 0.08  &                       0  &    $MW_{\text{fiducial}}$  &     $MW_{\text{fiducial}}$  &     600 &        0.16 $\pm$ 0.08  & 2.43 $\pm$ 0.74  &     1.66 $\pm$ 0.96 \\
BI-04 &                      $\pm$ 0.12  &                       0  &    $MW_{\text{fiducial}}$  &     $MW_{\text{fiducial}}$  &    600 &         0.24 $\pm$ 0.11  & 3.18 $\pm$ 0.99  &     2.34 $\pm$ 1.09 \\
\hline
BI-05 &                      $\pm$ 0.00  &                       0  &     $MW_{\text{fiducial}}$  &        $MW_{\text{low}}$  &     600 &        0.01 $\pm$ 0.03  & 1.07 $\pm$ 0.28  &     4.80 $\pm$ 1.50 \\
BI-06 &                      $\pm$ 0.04  &                       0  &     $MW_{\text{fiducial}}$  &        $MW_{\text{low}}$  &     600 &        0.05 $\pm$ 0.04  & 1.45 $\pm$ 0.34  &     5.19 $\pm$ 1.50 \\
BI-07 &                      $\pm$ 0.08  &                       0  &     $MW_{\text{fiducial}}$  &        $MW_{\text{low}}$  &     600 &        0.12 $\pm$ 0.07  & 2.11 $\pm$ 0.58  &     5.74 $\pm$ 1.54 \\
BI-08 &                      $\pm$ 0.12  &                       0  &     $MW_{\text{fiducial}}$  &        $MW_{\text{low}}$  &    600 &         0.20 $\pm$ 0.09  & 2.81 $\pm$ 0.8  &     6.15 $\pm$ 1.49 \\
\hline
BI-09 &                      $\pm$ 0.00  &                       0  &     $MW_{\text{fiducial}}$  &        $MW_{\text{high}}$  &    600 &         0.01 $\pm$ 0.02  & 1.05 $\pm$ 0.13  &     0.45 $\pm$ 0.50 \\
BI-10 &                      $\pm$ 0.04  &                       0  &    $MW_{\text{fiducial}}$  &         $MW_{\text{high}}$  &     600 &        0.07 $\pm$ 0.05  & 1.61 $\pm$ 0.42  &     0.52 $\pm$ 0.58 \\
BI-11 &                      $\pm$ 0.08  &                       0  &     $MW_{\text{fiducial}}$  &        $MW_{\text{high}}$  &     600 &        0.18 $\pm$ 0.09  & 2.58 $\pm$ 0.85  &     0.87 $\pm$ 0.77 \\
BI-12 &                      $\pm$ 0.12  &                       0  &     $MW_{\text{fiducial}}$  &        $MW_{\text{high}}$  &     600 &        0.25 $\pm$ 0.11  & 3.25 $\pm$ 0.95  &     1.39 $\pm$ 0.97 \\
\hline
BI-13 &                      $\pm$ 0.00  &                       5  &    $MW_{\text{fiducial}}$  &     $MW_{\text{fiducial}}$  &    600 &         0.02 $\pm$ 0.01  & 1.14 $\pm$ 0.12  &     1.35 $\pm$ 0.96 \\
BI-14 &                      $\pm$ 0.04  &                       5  &    $MW_{\text{fiducial}}$  &     $MW_{\text{fiducial}}$  &    600 &         0.07 $\pm$ 0.05  & 1.66 $\pm$ 0.42  &     1.53 $\pm$ 0.94 \\
BI-15 &                      $\pm$ 0.08  &                       5  &    $MW_{\text{fiducial}}$  &     $MW_{\text{fiducial}}$  &     600 &        0.17 $\pm$ 0.08  & 2.52 $\pm$ 0.76  &     2.28 $\pm$ 1.10 \\
BI-16 &                      $\pm$ 0.12  &                       5  &    $MW_{\text{fiducial}}$  &     $MW_{\text{fiducial}}$  &     600 &        0.24 $\pm$ 0.11  & 3.19 $\pm$ 0.96  &     2.98 $\pm$ 1.20 \\
\hline
BI-17 &                      $\pm$ 0.00  &                       5  &     $MW_{\text{fiducial}}$  &        $MW_{\text{low}}$  &      600 &       0.03 $\pm$ 0.04  & 1.26 $\pm$ 0.34  &     6.90 $\pm$ 1.76 \\
BI-18 &                      $\pm$ 0.04  &                       5  &   $MW_{\text{fiducial}}$  &          $MW_{\text{low}}$  &     600 &        0.06 $\pm$ 0.04  & 1.56 $\pm$ 0.36  &     7.38 $\pm$ 1.84 \\
BI-19 &                      $\pm$ 0.08  &                       5  &    $MW_{\text{fiducial}}$  &         $MW_{\text{low}}$  &      600 &       0.13 $\pm$ 0.06  & 2.15 $\pm$ 0.56  &     7.72 $\pm$ 1.86 \\
BI-20 &                      $\pm$ 0.12  &                       5  &     $MW_{\text{fiducial}}$  &        $MW_{\text{low}}$  &       600 &      0.21 $\pm$ 0.09  & 2.85 $\pm$ 0.81  &     8.09 $\pm$ 1.82 \\
\hline
BI-21 &                      $\pm$ 0.00  &                       5  &     $MW_{\text{fiducial}}$  &        $MW_{\text{high}}$  &     600 &        0.01 $\pm$ 0.01  & 1.09 $\pm$ 0.13  &     0.70 $\pm$ 0.64 \\
BI-22 &                      $\pm$ 0.04  &                       5  &     $MW_{\text{fiducial}}$  &        $MW_{\text{high}}$  &     600 &        0.08 $\pm$ 0.05  & 1.70 $\pm$ 0.46  &     0.76 $\pm$ 0.69 \\
BI-23 &                      $\pm$ 0.08  &                       5  &     $MW_{\text{fiducial}}$  &        $MW_{\text{high}}$  &       600 &      0.19 $\pm$ 0.09  & 2.67 $\pm$ 0.84  &     1.17 $\pm$ 0.84 \\
BI-24 &                      $\pm$ 0.12  &                       5  &     $MW_{\text{fiducial}}$  &        $MW_{\text{high}}$  &       600 &      0.26 $\pm$ 0.12  & 3.31 $\pm$ 1.07  &     1.82 $\pm$ 1.05 \\

\hline
\end{tabular}
\end{table*}

\section{Additional checks}
\label{AppendixPart2}
Apart from the main results, we also checked some additional checks to see the behavior of the satellite plane. 

\subsection{Correlated PM and Dist errors}
\label{AppendixPart2_recomputing_corr}

\begin{figure*}
\centering
   \includegraphics[width=15cm]{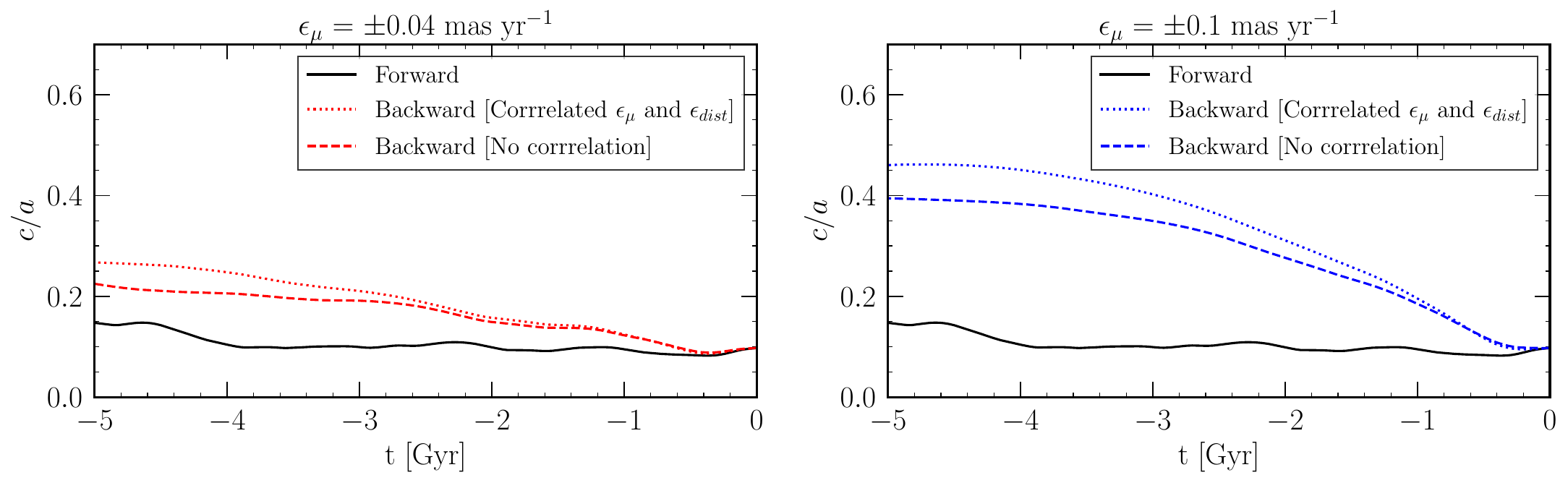}
     \caption{Comparison of correlated and uncorrelated PM-distance errors on satellite plane evolution. Two cases are shown with PM errors of 0.04 and 0.1 mas yr$^{-1}$, both with a 10\% distance error. Correlated errors lead to a slightly increased plane widening.}
     \label{fig:plot_ca_dist_pm_err_corr}
\end{figure*}

We also test whether a correlation between proper motion and distance errors affects the inferred evolution of the satellite plane, as fainter test satellites could have both larger proper motion and distance uncertainties. To investigate this, we performed an additional analysis, comparing two cases (see Fig. \ref{fig:plot_ca_dist_pm_err_corr}): one with a proper motion error of 0.04 mas yr\(^{-1}\) and a 10\% distance error, and another with a PM error of 0.1 mas yr\(^{-1}\) and a 10\% distance error. Each plot includes two backward integrations—one with a PM-distance error correlation and one without. In both cases, the same errors are used, but in the 'correlated' case, the errors are ranked and assigned to the test satellites such that satellites farther away have progressively larger distance and proper motion errors. We find that the effect is not extreme but leads to an increase in the inferred widening of the satellite plane. This suggests that the approach of using uncorrelated errors is conservative and at most underestimates the impact of uncertainties. Moreover, as expected, with higher PM error, the effect becomes more pronounced, aligning with what we observed in our main results.

\subsection{Backward integration with \texttt{MilkyWayPotential2022}}
\label{AppendixPart2_recomputing_gala}

\begin{figure*}
\centering
   \includegraphics[width=13cm]{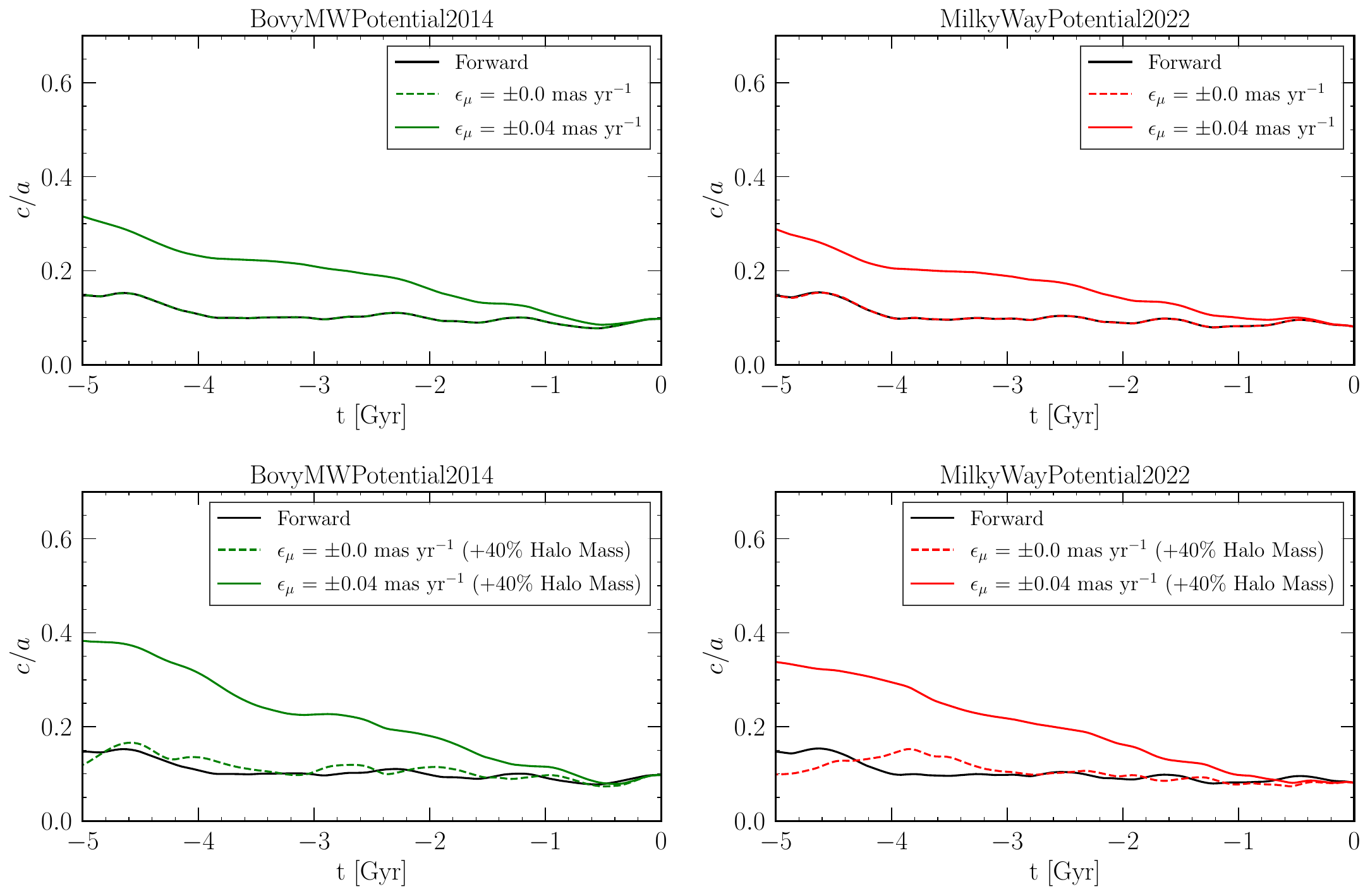}
     \caption{Comparison between \texttt{MilkyWayPotential2022}  and \texttt{MWPotential2014} }
     \label{fig:GalpyGalaPot}
\end{figure*}

In Fig. \ref{fig:GalpyGalaPot}, we re-ran some of our simulations and show a comparison between the backward integration done by \texttt{MilkyWayPotential2022} from \texttt{Gala}. We found that our results remain qualitatively unchanged. In Fig. \ref{fig:GalpyGalaPot}, the first column corresponds to \texttt{MWPotential2014}, and the second to \texttt{MilkyWayPotential2022}. In each plot, we include one example with no proper motion error and one with a 0.04 mas/yr proper motion error. While some minor differences are noticeable, the overall behavior and results of our study remain unaffected by the choice of potential. We also tested a +40\% halo mass variation for both \texttt{MWPotential2014} and \texttt{MilkyWayPotential2022}, and the results remained consistent.

\subsection{Re-computing $c/a$ but with test
satellites within 300 kpc}
\label{AppendixPart2_recomputingc_a_300}

\begin{figure*}
\centering
   \includegraphics[width=13cm]{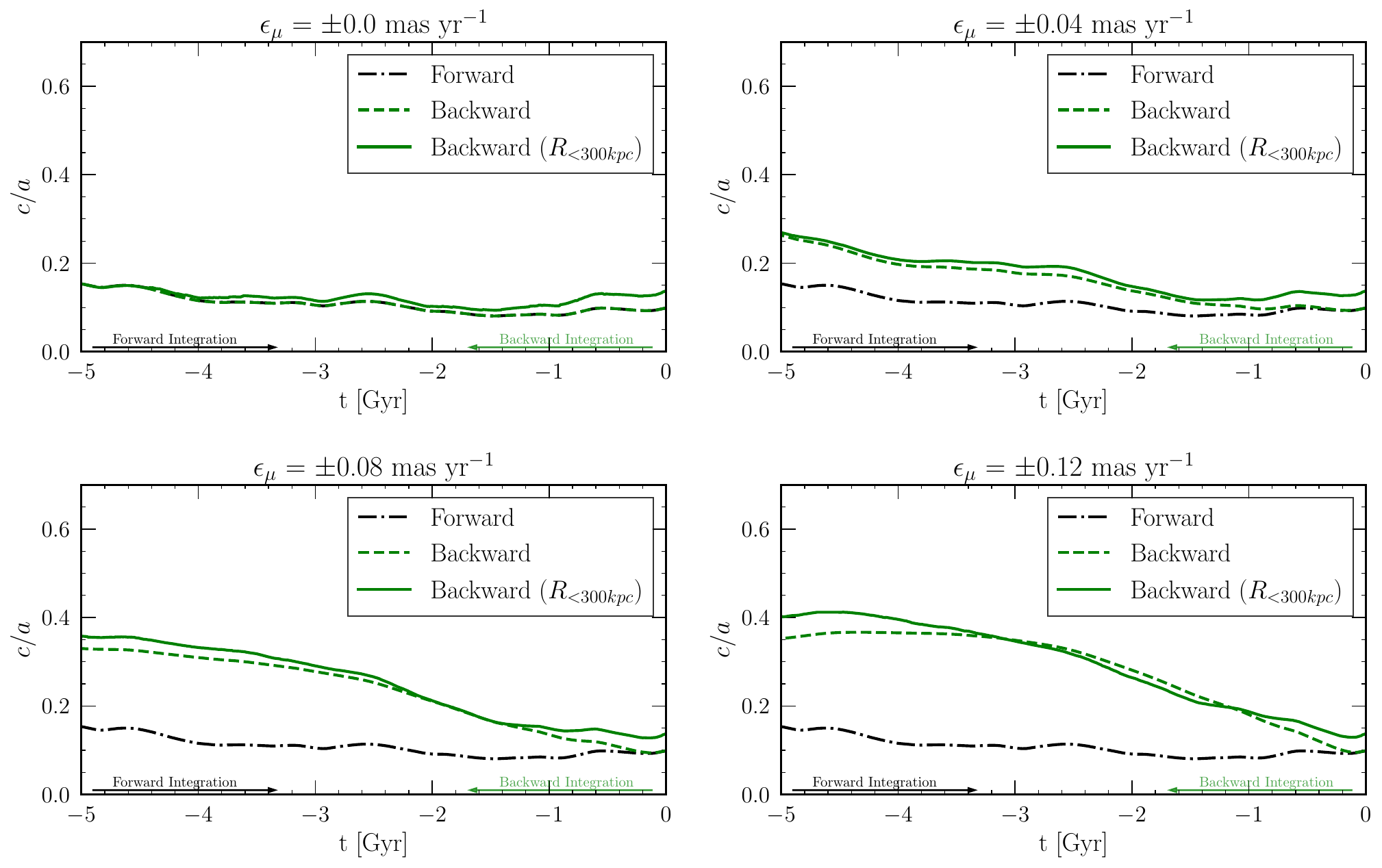}
     \caption{Evolution of \( c/a \) for the \( MW_{\text{fiducial}} \) model with and without excluding test satellites beyond 300 kpc. Dashed and solid green lines show backward evolution, black lines indicate forward integration. Exclusion has minimal impact on the overall trend.}
     \label{fig:ca80_no_dist_less300}
\end{figure*}

Figure \ref{fig:ca80_no_dist_less300} shows the evolution of \( c/a \) for the \( MW_{\text{fiducial}} \) potential model, comparing cases with and without the exclusion of test satellites beyond 300 kpc. The dashed green line represents the backward evolution when these satellites are excluded, while the solid green line shows the backward evolution including all test particles. The black lines indicate the forward integration. The exclusion of satellites beyond 300 kpc does not significantly alter the overall evolution of \( c/a \). While removing some test satellites has a minor effect on the backward integration of \( c/a \), this effect is not substantial if satellites beyond 300 kpc are excluded.

\end{appendix}

\end{document}